# Ride, Track, and Recover: Pilot Randomized Trial of a Wearable Digital Self-Management Intervention During a Veteran Endurance-Cycling Program

Alan Ta[a] (alanta2002@tamu.edu), Nilsu Salgin[a] (nilsusalgin@tamu.edu), Caleb Armstrong[a] (amhd@tamu.edu), Kala Phillips Reindel[b] (kpr@tamu.edu), and Farzan Sasangohar [a]* (sasangohar@tamu.edu)

[a]Department of Industrial and Systems Engineering, Texas A&M University, 3127 TAMU, College Station, TX 77843-3127

[b]Texas A&M Health Telehealth Institute, 1700 Research Pkwy, Suite 250, College Station, TX 77845

*Corresponding Author: Farzan Sasangohar (email: sasangohar@tamu.edu)

Wm Michael Barnes ’64 Department of Industrial & Systems Engineering

Texas A&M University

Emerging Technologies Building

College Station, TX 77843-3131

**ABSTRACT**

**Introduction**: Post-traumatic stress disorder (PTSD) in veterans is characterized by persistent hyperarousal and comorbid anxiety and depressive symptoms that are difficult to monitor and manage outside clinical settings.

**Objective**: This pilot study evaluated a wearable-integrated digital mental health system embedded within a real-world endurance cycling intervention to examine longitudinal hyperarousal dynamics, symptom trajectories, and user-centered performance of a machine learning (ML)–based detection system.

**Methods**: Thirteen veterans participating in a Project Hero cycling event in Texas were randomized by computer-generated sequence in a naturalistic setting to two arms: (1) digital intervention plus physical activity, or (2) physical activity only, plus a third at-home monitoring control cohort consisting of 7 veterans selected from the broader Project Hero veteran community. Continuous smartwatch sensing combined heart rate and accelerometer features to detect hyperarousal events, which were confirmed in real time by participants. Weekly self-report measures of anxiety (GAD-7), depression (PHQ-8), and PTSD severity (PCL-5) were collected. Generalized additive mixed models (GAMMs) characterized nonlinear trajectories over time (Clinical Trial Number: NCT06993012).

**Results**: Baseline-normalized hyperarousal trajectories differed significantly across conditions, with the digital intervention group (n=7) showing structured stabilization compared to late-study escalation in the physical-only group (n=3). Both cycling groups exhibited acute symptom improvements during the endurance event; however, the digital intervention group demonstrated a higher overall maintenance of gains. Theat-home control group (n=4) showed gradual symptom declines. Perceived precision of ML detections varied substantially across individuals and was positively associated with symptom severity, with higher-severity participants confirming a greater proportion of detected events. Qualitative findings indicated that real-time detection increased awareness and mindfulness but highlighted usability barriers and differing expectations for post-alert support.

**Conclusion**: Together, these results suggest that coupling wearable detection with digital self-management tools may support stabilization of hyperarousal and symptom improvement while emphasizing the importance of personalization and human-centered design in wearable mental health systems.

# 1. INTRODUCTION

Post-Traumatic Stress Disorder (PTSD) arises after experiencing or witnessing a life-threatening event, such as trauma experienced from combat, near-death injury, or witnessing violence. It involves four symptom clusters: intrusive recollections, hyperarousal, avoidance, and negative mood/cognitions, all of which interfere with everyday living (Mann et al., 2025).

Impacts of PTSD go beyond psychological morbidity, with severe impacts across the lifespan. It interferes with quality of life through compromised occupational functioning, difficulty in social and family relationships, and decline in physical health. In addition, PTSD frequently co-occurs with other diseases such as depression, alcohol use disorders, and cardiovascular illnesses that further add to the shared load of PTSD on healthcare systems and patients (Miller et al., 2024). Comorbidity underscores the multi-dimensional nature of the challenges generated by PTSD and its hybrid needs for treatment.

Prevalence and severity of PTSD are significantly greater among military groups, especially active combat veterans, than civilians. Seminal research of Operation Iraqi Freedom (OIF) and Operation Enduring Freedom (OEF) veterans found that 25% had at least one mental illness diagnosis (Seal et al., 2007). Among these, 44% reported a single diagnosis, 29% two diagnoses, and 27% three or more, which illustrates the high degree of psychiatric comorbidity among this population. Notably, PTSD was the most common disorder, and 13% of the 102,632 veterans researched had PTSD (Seal et al., 2007). The results reflect the prevalence and complexity of PTSD among returning troops.

## 1.1. Traditional Care Models and Limitations

Evidence-based treatments for PTSD include trauma-focused psychotherapies such as cognitive processing therapy and prolonged exposure, as well as pharmacological interventions including selective serotonin reuptake inhibitors, serotonin-norepinephrine reuptake inhibitors, and adjunctive medications (Schrader & Ross, 2021). While these approaches are effective for many individuals, real-world implementation is constrained by substantial barriers.

Veterans often face stigma related to mental health treatment, limited availability of specialty services, and logistical challenges such as travel, scheduling, and cost. These barriers contribute to high dropout rates and inconsistent engagement, particularly after structured treatment programs end (Amsalem et al., 2022). Moreover, PTSD symptoms frequently recur in the absence of long-term monitoring and support, underscoring the limitations of traditional care models (Brooks & Greenberg, 2024). Together, these challenges highlight the need for scalable, continuous, and self-directed interventions that can complement traditional clinical services.

Group-based physical activity and social engagement programs have emerged as promising adjunctive approaches for veterans with PTSD. Programs such as Project Hero integrate endurance cycling, peer support, and community engagement to promote resilience, reduce stress, and foster social connectedness (Schnor et al., 2019). Participation in such programs has been associated with improvements in mood, physical health, and perceived social support (Ma & Mumtaz, 2025).

However, these benefits often diminish once programs conclude, and sustaining long-term engagement remains difficult if not enrolled in a structured program (Austin et al., 2025). In addition, in-person programs are resource-intensive and challenging to scale across geographically dispersed veteran populations. As a result, there is a critical need for tools that can extend the benefits of these interventions beyond the event itself and into veterans' daily lives.

### 1.2. Mobile Health (mHealth) for PTSD Support

Mobile health (mHealth) refers to the use of smartphones, wearable devices, and connected applications to support health monitoring, intervention delivery, and self-management (Marcolino et al., 2018). For PTSD, mHealth platforms offer unique advantages, including continuous monitoring, personalized feedback, and just-in-time interventions delivered outside of traditional clinical settings (Rodriguez-Paras et al., 2017).

For veterans, mHealth tools may reduce barriers related to stigma and access by providing discreet, self-guided support while extending the reach of in-person programs (Y. Yang et al., 2022). When integrated with wearable sensors, these platforms enable passive, real-time data collection that can inform adaptive interventions tailored to individual needs.

Wearable accelerometers, commonly embedded in smartwatches, can capture movement patterns and physiological data that correlate with stress. These signals can be used to infer physical activity, arousal states, and stress-related behaviors (Lazarou & Exarchos, 2024; Napolitano et al., 2010). When paired with heart rate and heart rate variability data, accelerometer features can support machine learning (ML) models that detect stress in real time (Hongn et al., 2025).

These objective, passive sensing approaches offer a means of identifying fluctuations in stress and self-regulation capacity without relying solely on self-reporting, which is often burdensome and prone to recall bias. This capability is particularly relevant for hyperarousal, one of the core symptom clusters of PTSD, which manifests through both physiological and behavioral changes.

### 1.3. Prior Research

Advances in wearable sensing technologies, particularly smartwatches, have enabled continuous, nonintrusive collection of physiological data in real-world settings. Prior work has shown that several physiological measures, including heart rate and heart rate variability, are associated with PTSD symptomatology, with heart rate emerging as a reliable correlate of hyperarousal states (McDonald et al., 2019; Sadeghi et al., 2022). Building on this foundation, machine learning models have been developed to detect PTSD-related hyperarousal events in real time using wearable sensor data, with promising classification performance in combat veteran populations.

However, existing studies have focused on computational or laboratory-based validation, with limited attention to naturalistic deployment and user-centered evaluation (Mishra et al., 2020). Research has demonstrated substantial variability in perceived precision of automated hyperarousal detection, highlighting the importance of aligning physiological detection with users' subjective experiences to support trust and sustained engagement (Sadeghi et al., 2021).

The present study extends existing work by evaluating a wearable-integrated digital mental health system, the First Watch Device (FWD) platform, in a naturalistic, longitudinal context involving veterans participating in an intensive endurance cycling intervention. The FWD platform serves as an accessible, continuous monitoring system for increasing users' awareness of real-time physiological stress responses. By providing immediate feedback on hyperarousal events, FWD is designed to enable veterans to recognize stress patterns as they occur and engage in self-management strategies. Hyperarousal is confirmed using a hybrid metric, Moments of Stress (MS), which combines machine-learning-based physiological detection with real-time user confirmation (Ta et al., 2025). In this study, real-time hyperarousal events were identified using a newly developed machine-learning model that integrates heart-rate and accelerometer-based features. This model builds on heart-rate-driven approaches by incorporating multimodal wearable signals to improve the detection of stress-related physiological states, as originally validated in a naturalistic veteran population (Sadeghi et al., 2022).

## 2. CURRENT STUDY

### 2.1. Experimental Design

This study investigates the feasibility and preliminary efficacy of a wearable-integrated digital mental health intervention designed to support veteran-centered PTSD symptom monitoring and self-management during and surrounding an intensive endurance cycling event. Specifically, the study examines changes in real-time hyperarousal events, as well as self-reported PTSD, anxiety, and depressive symptoms, among veterans participating in Project Hero's Texas Challenge and a control group of veterans (Project Hero, 2025).

The study employed a single-blind, three-group randomized controlled design conducted in a naturalistic setting. Participants were assigned to one of three conditions: (1) a combined digital health self-management intervention and group-based physical activity intervention, (2) a group-based physical activity–only intervention (active comparator), and (3) an at-home control condition drawn from the broader Project Hero veteran community not registered for the Texas Challenge event. Randomization for the two intervention arms was performed using a computer-generated sequence, while at-home control participants were selected from eligible veterans who did not participate in the endurance cycling event. The at-home control cohort was not formally matched to the cycling arms on demographic or clinical characteristics; given the pilot nature of the study, group comparability was assessed descriptively rather than through stratified randomization.

Mental health outcomes were assessed longitudinally across three primary phases: pre-intervention, during the intervention, and post-intervention, spanning from April 23, 2025, to June 7, 2025. The physical intervention phase consisted of the 7-day Texas Challenge, during which participants in the two intervention arms cycled approximately 8 hours per day along a predetermined route. The Texas Challenge took place from May 4-10, 2025, with a route from San Antonio, Texas, moving north, and ending in Arlington, Texas. All participants wore an Apple Watch and used a paired smartphone application throughout their participation to enable continuous physiological monitoring and repeated psychological assessments. The research team reviewed data daily to ensure data quality and participant safety, and outcome assessors were blinded to group assignments.

Project Hero leadership was consulted during study planning to assess the feasibility of wearable deployment during endurance cycling. Feedback-informed onboarding procedures and selection of self-report instruments. Project Hero was not involved in data analysis or manuscript preparation.

## 2.2. Research Questions

This study examined four primary research aims related to the effects and performance of a wearable-enabled PTSD self-management system during a real-world endurance cycling intervention. The aims address: (1a) Hyperarousal Trajectories, (1b) Psychological Symptom Trajectories, (2a) Perceived Precision of the Detection Model, and (2b) Veteran Perspectives and Takeaways.

**RQ1(a):** Do veterans receiving digital self-management tools in addition to wearable monitoring demonstrate different longitudinal hyperarousal trajectories compared to veterans receiving wearable monitoring alone? We hypothesized (H1a) that extended use of the full suite of self-management features within the FWD system would be associated with a decreased frequency of identified MS across the study period, reflecting improved regulation of acute hyperarousal. Increased situational awareness, combined with just-in-time access to self-management tools, is expected to facilitate more effective coping with acute stressors. In contrast, participants who do not receive such real-time feedback and digital support are expected to show smaller or more variable changes in hyperarousal frequency, driven primarily by contextual fluctuations rather than intervention effects.

**RQ1(b):** Do veterans receiving digital self-management tools demonstrate different longitudinal changes in PTSD, anxiety, and depressive symptoms compared to monitoring-only conditions? Following similar reasoning, we hypothesized (H1b) that extended use of the full suite of self-management features within the FWD would be associated with longitudinal improvements in self-reported PTSD, anxiety, and depressive symptoms. The FWD system serves as a digital self-management and monitoring tool to help veterans understand their psychological and physiological stress responses. Increased awareness of hyperarousal, combined with opportunities to engage in adaptive coping strategies and reflective self-monitoring, is expected to promote better emotional regulation and symptom management over time.

**RQ2(a):** To what extent does the ML-based hyperarousal detection model demonstrate acceptable perceived precision in real-world use? We hypothesized (H2a) that the FWD smartwatch detection system would demonstrate moderate to high perceived precision, regardless of study condition, with the majority of automated hyperarousal detections being confirmed by users. Research on wearable-based PTSD detection has shown that heart-rate-driven machine learning models can reliably capture physiological patterns associated with hyperarousal (Sadeghi et al., 2022). By integrating physiological sensing and prompting users to confirm detected events in real time, the FWD system is expected to achieve acceptable alignment between automated detection and subjective experience.

**RQ2(b):** How do veterans in the digital + physical intervention group perceive the usability, trustworthiness, and overall user experience of the FWD system in supporting PTSD self-management? We hypothesized (H2b) that veterans would report positive perceptions of the full FWD system's usability and trust. The FWD platform was designed using human factors and user-centered design principles, with an emphasis on minimal interaction burden, simple and intuitive interfaces, and seamless integration into daily life via smartphones and smartwatches. Real-time notifications, simple confirmation interactions, and passive data collection are intended to reduce cognitive and behavioral workload while maintaining meaningful engagement.

## 3. METHODOLOGY

### 3.1. Power Analysis and Participants

An a priori power analysis was conducted in G*Power 3.1 (Faul et al., 2007) for a repeated-measures ANOVA of the within–between interaction (group × time) with three groups and seven measurement occasions. Assuming $\alpha = .05$, 80% power, a medium effect ($\eta_p^2 = .06$; Cohen's $f = 0.253$; (Cohen, 1988)), an average within-subject correlation of $r = .45$, and $\varepsilon = .90$, the required sample size was $N = 27$ (≈ 9 per group; achieved power ≈ .84). To account for 20% attrition, the recruitment target was set at $N = 36$ (12 per group).

20 U.S. military veterans were enrolled through the Project Hero community. Eligibility criteria included age ≥ 18 years, self-identified U.S. military veteran status, willingness to wear a smartwatch and use a mobile health application, and access to a smartphone with reliable internet. Additionally, participants completed the PTSD Checklist for DSM-5 (PCL-5) prior to enrollment, with individuals scoring > 10 being eligible for study inclusion. Exclusion criteria included inability to provide informed consent, age < 18 years, pregnancy when participation could pose a risk, and incarceration. 13 of the 20 participants were randomly assigned to two intervention groups. Participants in the intervention conditions were recruited from registrants of Project Hero's 7-day Texas Challenge (Project Hero, 2025). The other 7 participants were recruited from the broader Project Hero community who were not registered for the Texas Challenge. Attrition due to scheduling constraints and technology issues reduced the analyzable sample to 14 participants ($M_{age} = 56.8$, $SD_{Age} = 6.95$) with digital + physical intervention: $n$ =

7; physical intervention only: $n$ = 3; at-home control: $n$ = 4; therefore, outcomes are interpreted as exploratory/pilot estimates of feasibility and effect patterns for future adequately powered trials. 7 of the 14 participants were Caucasian, 3 were Hispanic, 2 were African American, 1 was Asian, and 1 was of undisclosed ethnicity. A more detailed distribution of participants across the study arms and reasons for study attrition can be seen in **Figure 1** according to the CONSORT 2025 flow diagram (Hopewell et al., 2025).

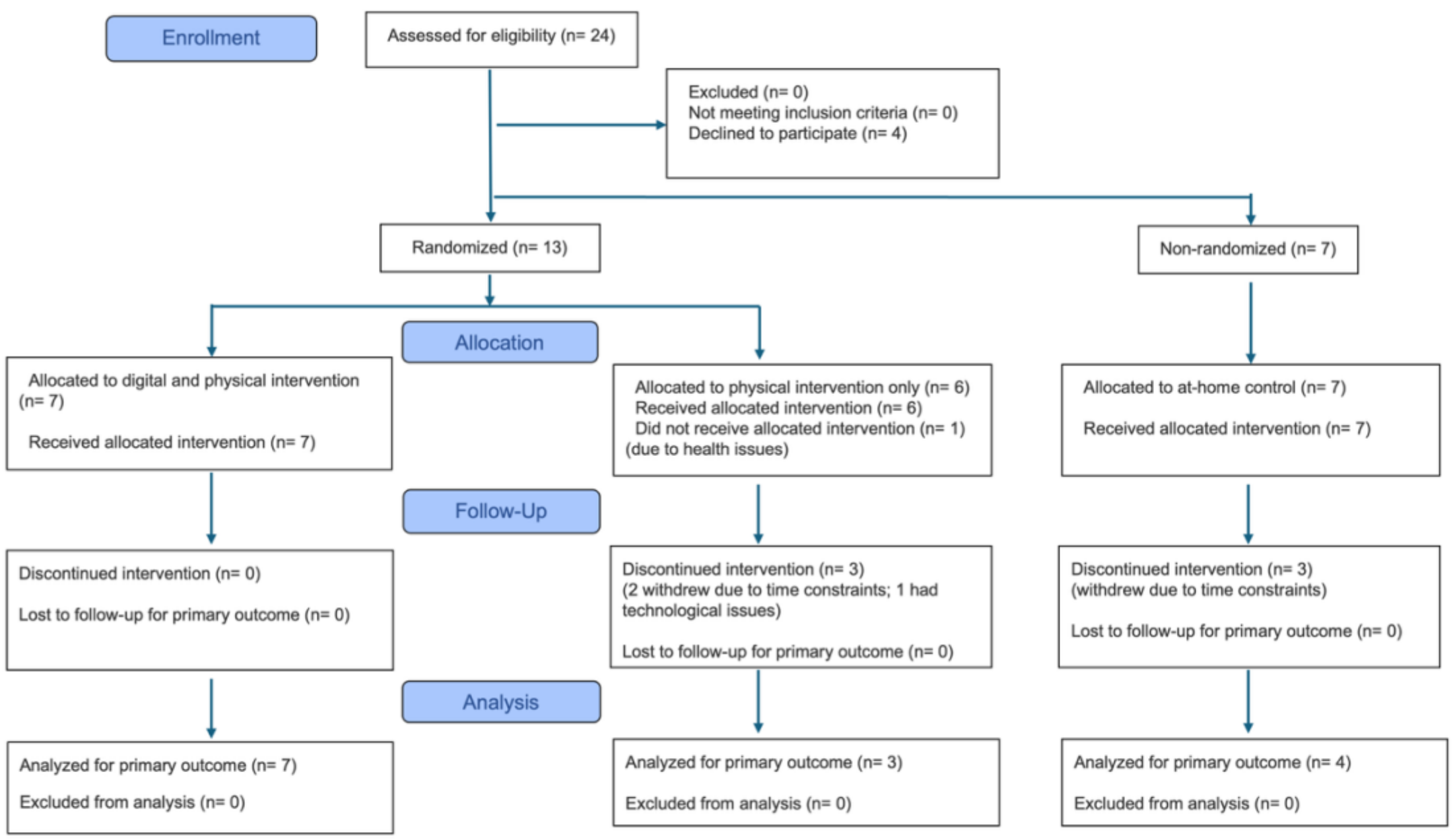

**Figure 1:** CONSORT 2025 Flow Diagram. The three columns represent the 3 arms of the studies. Participant distribution and reasons for study withdrawal or loss of follow-up are depicted.

### 3.2. First Watch Device Platform

The study used the First Watch Device (FWD), a smartphone application paired with a smartwatch interface that functions as an integrated platform for mental health symptom monitoring and self-management among veterans. The FWD mobile application brings key functions together in a single interface, including self-management supports, psychoeducational materials, safety and resource features, contact management, and symptom and behavior tracking views (**Figure 2**). The Home tab provides a day-level overview and a task-oriented workflow that prompts completion of scheduled self-assessments (e.g., PCL-5 and GAD-7) via the task list. App navigation is organized through five bottom-bar tabs. The Engage tab contains self-management tools (e.g., relaxation media, guided breathing, focus supports, motivational content, and journaling) and includes an embedded Education section. The Education section categorizes content by symptom-relevant topics (e.g., depression, stress, anxiety, PTSD) and presents curated videos and articles. For acute needs, the SOS tab provides rapid access to crisis and emergency actions, including direct connection to the Veterans Crisis Line, messaging options, and a feature for locating nearby emergency rooms. The Link tab supports organization of a personal support network through groups (e.g., family), care provider entries, and an all-contacts list. The Progress tab presents available tracking tiles, including heart rate, stress moments, sleep tracking, self-assessments, and medication.

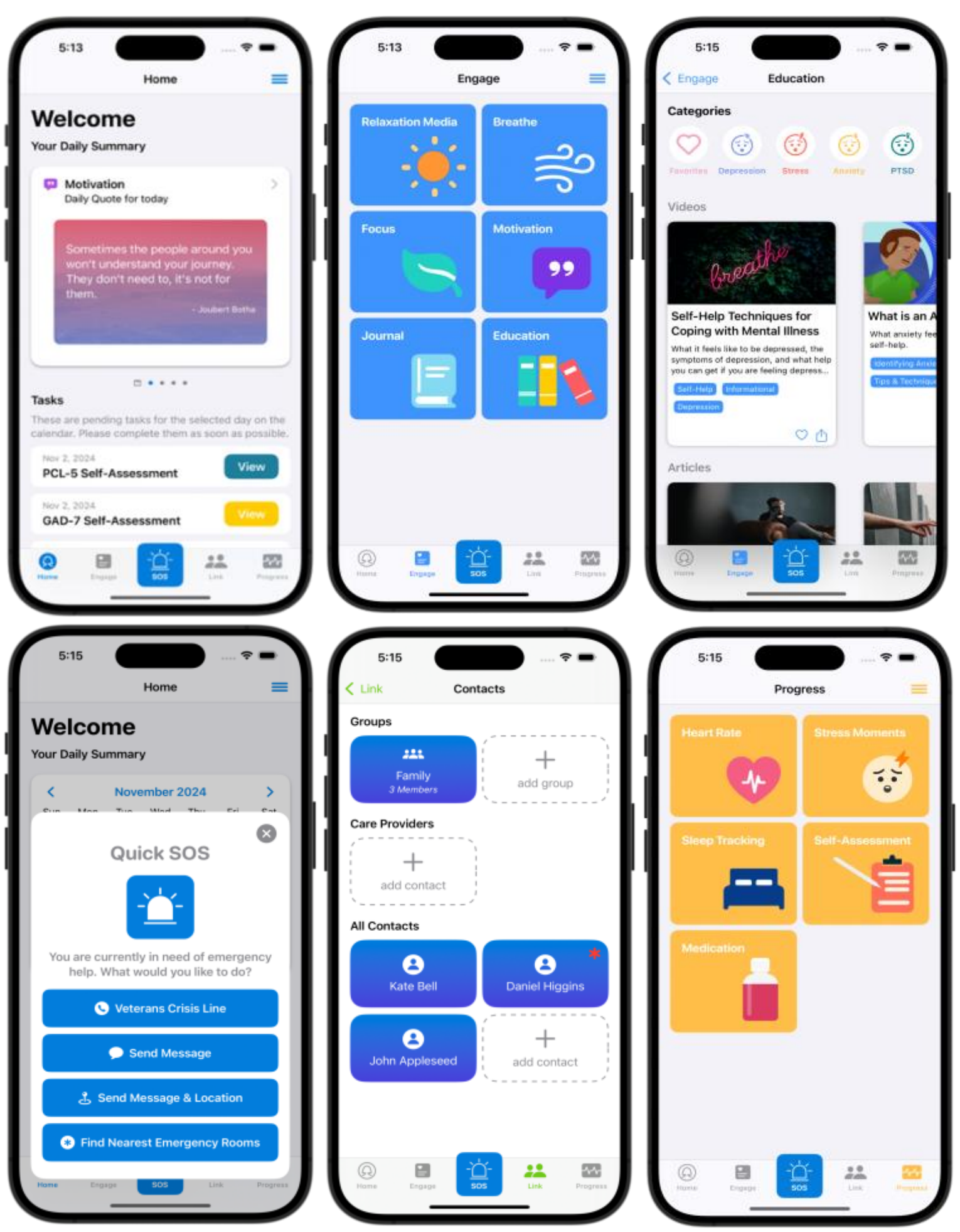

***Figure* 2.** Sequence of core FWD smartphone navigation screens. From left to right, the app displays: (1) the home dashboard with calendar and assessment tasks, (2) the Engage tab offering stress management tools (e.g., breathing, journaling, motivation), (3) the Education feature for helpful videos and articles regarding mental health conditions, (4) the Quick SOS feature for sending emergency messages to contacts, (5) the Link tab with personal contacts and emergency resources, and the (6) the Progress tab tracking health metrics (e.g., heart rate, stress moments, sleep).

The FWD smartwatch application (**Figure 3**) is paired with the smartphone app and provides a wrist-based interface for interacting with the system. The watch app displays notifications when the machine-learning algorithm detects elevated stress, asking the user whether to record a stress moment; responses are captured via Yes/No selections. The smartwatch interface also offers quick access to core on-device options, including Start Monitoring, Stress Moment, Heart Rate, and Settings. Use of the intervention features was not standardized; instead, participants were encouraged to engage with available resources in a manner that aligned with their individual needs. Accordingly, FWD was treated as a bundled, multi-component digital mental health intervention, and the trial assessed the overall effect of the integrated system rather than attributing outcomes to any single element.

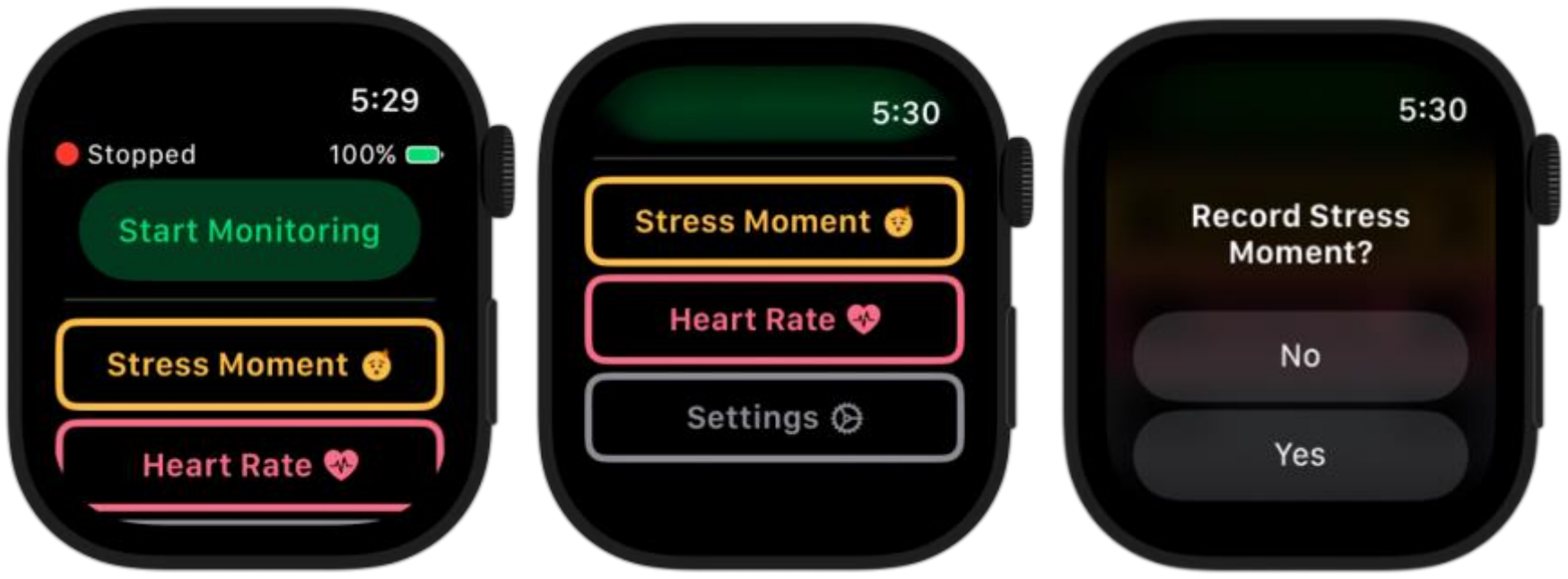

**Figure 3.** Sequence of FWD Apple Watch interface screens. The smartwatch interface allows users to: (1) start/stop monitoring, report stress moments, and view heart rate, (2) manage data sharing settings, and (3) respond to detected stress prompts.

### 3.3. Procedure

After providing informed consent, eligible participants were enrolled on a rolling basis and assigned to their respective study conditions. Due to staggered enrollment and event-based constraints, participants initiated study monitoring at varying time points, with a maximum pre-intervention monitoring period of approximately two weeks. Some participants began physiological monitoring and psychological assessment at the start of the Texas Challenge, while others collected pre-event baseline data.

The study followed a structured, multi-phase timeline spanning eight weeks, comprising a continuous monitoring period and a mid-study physical intervention (**Figure 4**). Throughout the study, all participants wore the smartwatch with hyperarousal detection enabled and completed weekly behavioral health surveys, enabling continuous collection of physiological, behavioral, and self-reported psychological data across all study arms. Participants in the digital-plus-physical intervention group had access to the full FWD platform, including real-time hyperarousal detection and digital self-management tools. In contrast, participants in the physical intervention-only group had access to hyperarousal detection and monitoring features but did not receive digital self-management support. Participants in the at-home control group did not participate in the cycling event and had access only to wearable-based detection/monitoring and weekly behavioral surveys.

All participants were instructed to wear an Apple Watch continuously throughout their period of participation, except during activities such as charging or bathing. The smartwatch was paired with a smartphone application that served as the central platform for collecting physiological data, reporting hyperarousal events, and conducting psychological assessments. When the wearable-based monitoring system detected a potential hyperarousal event, participants across all study arms were prompted to confirm the experience and were encouraged to engage in appropriate self-management strategies.

During Week 1 (pre-event monitoring), baseline physiological and behavioral measurements were established under normal daily conditions. The cycling intervention occurred in Week 3, during which participants assigned to the cycling groups participated in a multi-day endurance cycling challenge while continuing to undergo wearable-based monitoring and hyperarousal detection. Participants in the at-home control group did not engage in the cycling event but remained fully instrumented. Following the cycling challenge, participants were monitored during a post-event phase in Week 4 and through the end of the study in Week 8, enabling longitudinal assessment of both short-term and sustained changes in hyperarousal patterns and behavioral health outcomes.

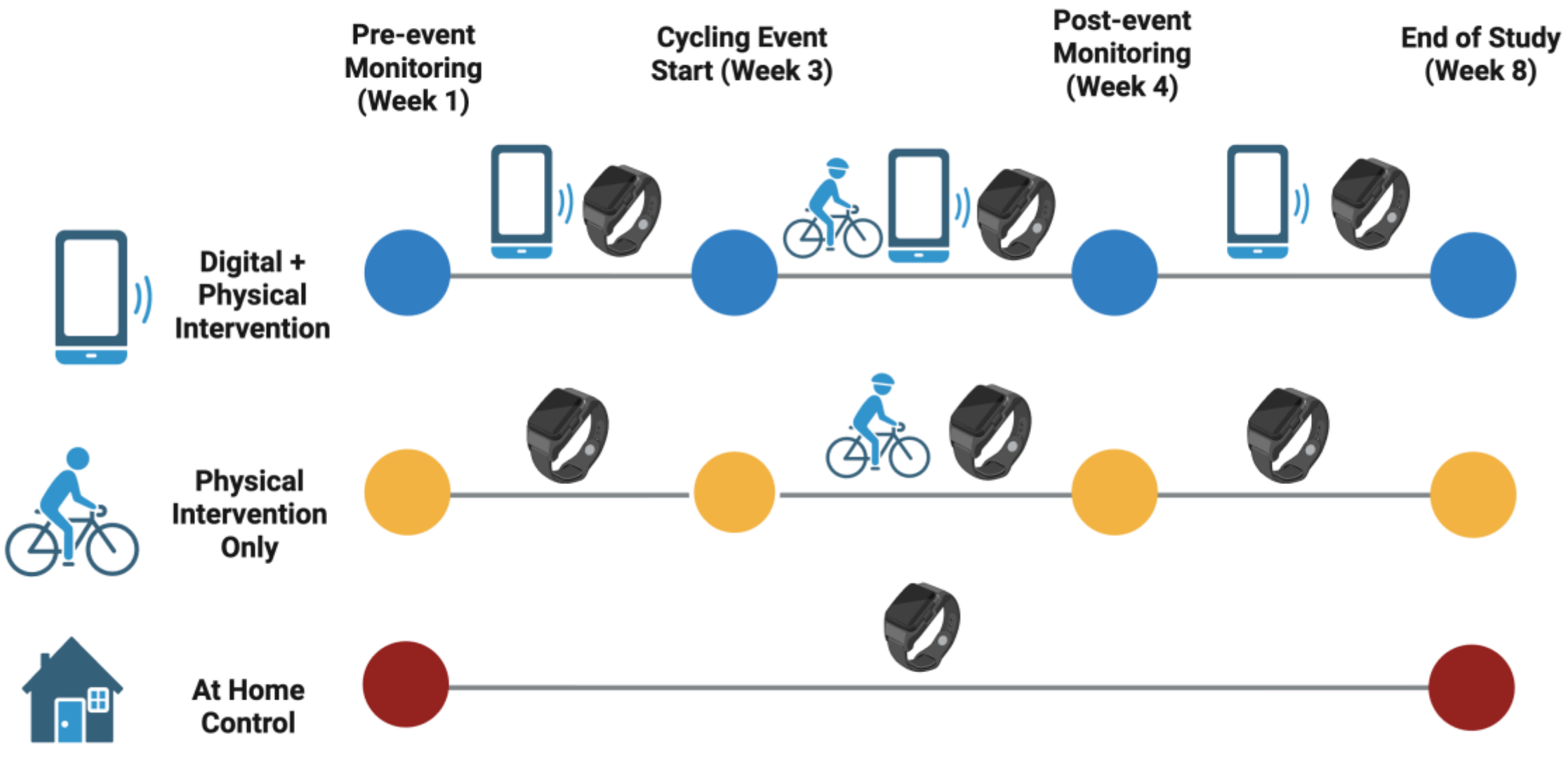

**Figure 4.** Experimental timeline across the three arms, created in BioRender. (https://BioRender.com/1at3fbd) is licensed under CC BY 4.0.

### 3.4. Measures

The primary dependent variable, Moments of Stress (MS), is a conceptual, real-time measure derived from a combination of physiological heart rate (beats per minute [bpm]) and participant self-reports, both collected via the Apple Watch smartwatch application. Participants wore an Apple Watch Series 4 or 5 with the app installed throughout the study. The app ran continuously in the background and connected to participants' phones to transfer data. The app continuously and remotely collected physiological and activity data, including heart rate and acceleration, at 1 Hz, and allowed participants to self-report a PTSD hyperarousal event or "Moment of Stress" via a double-tap on the watch face, generating a time-stamped self-report (Sadeghi et al., 2022).

In addition to participant-reported events, MS were identified from physiological time-series data using an integrated detection approach based on heart rate dynamics. Specifically, the algorithm analyzed HR fluctuations and automatically logged detected events in the exported dataset under a designated detection field. These real-time hyperarousal events were identified using an extended multimodal machine learning model that builds upon prior heart-rate–based validation work (Sadeghi et al., 2022). To operationalize MS for analysis, events were coded in a binary manner: a value of '1' was assigned when an algorithm-detected event was subsequently confirmed by the participant, whereas '0' represented time periods without detected or self-reported hyperarousal. Participant-reported MS were retained with their time-stamps to complement sensor-driven detection and provide additional context. Both algorithm-detected/confirmed and self-reported events were aggregated daily to derive individual MS frequencies, reflecting the total number of hyperarousal events recorded for each participant per day.

An additional metric, perceived precision, was defined as the proportion of automated hyperarousal detections from the ML-based smartwatch app that participants confirmed. Specifically, perceived precision was computed as the ratio of true positive detections to the total number of system-detected events, depicted in Formula 1, such that:

$$Perceived\ Precision = \frac{Number\ of\ True\ Positives}{(Number\ of\ True\ Positives + Number\ of\ False\ Positives)} \quad (1)$$

where a true positive corresponded to a system-detected event for which the participant responded "Yes" to the notification, and a false positive corresponded to a system-detected event for which the participant responded "No". These were calculated on a daily and weekly basis for each participant across all arms.

Three self-administered psychological assessments were collected as secondary outcome measures: the Generalized Anxiety Disorder 7-item (GAD-7), Patient Health Questionnaire 8-item (PHQ-8), and PTSD Checklist for DSM-5 (PCL-5). The GAD-7 is a brief questionnaire that asks individuals how often, within the last two weeks, they experience various symptoms of anxiety from a range of "not at all" to "nearly every day" (Spitzer et al., 2006). The PHQ-8 is a similarly designed survey that asks individuals about depressive symptoms (Kroenke et al., 2009). The GAD-7 and PHQ-8 have maximum scores of 21 and 24, respectively, with higher scores indicative of greater severity of the corresponding mental health condition. The PCL-5 is a 20-item self-report measure that assesses DSM-5 symptoms of PTSD. The PCL-5 has a maximum score of 80, with higher scores indicating greater symptom severity, typically asking about patient symptoms within the past month (Weathers et al., 2013). Within this study, versions of the GAD-7, PHQ-8, and PCL-5 that asked about symptom severity on a weekly basis ("in the past week…") in order to increase sensitivity to week-by-week changes in condition. In the case of severe distress, participants were referred to appropriate mental health resources, including veteran crisis lines and emergency services, in accordance with the approved data monitoring plan.

Once a week, the study application prompted participants to complete these psychological surveys. If participants were unresponsive or behind on their study tasks, the app sent daily reminder notifications to encourage completion. Survey responses, alongside MS records, were stored in secure cloud-based study storage (e.g., AWS buckets), with exact completion timestamps preserved in the raw export. Members of the research team monitored incoming data daily to ensure that physiological and survey data were being collected in a timely manner and to support follow-up reminders as needed. Surveys that were not completed were retained as missing (no imputation), and longitudinal analyses used all available completed survey observations.

To extend our quantitative findings, we conducted a structured post-study debrief through an exit survey delivered to all participants as a knowledge-elicitation procedure (Cooke, 1994). This approach enabled us to systematically capture participants' cognitive interpretations of detection alerts and the specific actions they undertook in response. Two coders independently applied thematic analysis to the open-ended responses, thereby identifying recurring patterns in cue recognition, interpretation, and subsequent action, as well as barriers to adoption. The codes were then refined into one set of themes through consensus between the two coders. Accordingly, participants were asked to complete exit surveys (i.e., post-study debrief; N=12) that included questions about their experience with the detection tool embedded in FWD.

## 4. DATA ANALYTICS AND RESULTS

Generalized Additive Mixed Models (GAMMs; Pedersen et al., 2019) were applied to characterize longitudinal change in three complementary outcomes: Moment of Stress, weekly symptom self-assessment scores (GAD-7, PCL-5, PHQ-8), and perceived precision. GAMMs were used because the time course was not assumed to be linear and could differ across conditions. Time (day or week) was represented by flexible nonlinear functions, estimated separately for each condition, so that the shape of change could vary across conditions without imposing a specific functional form (e.g., linear or quadratic). For both MS, symptom outcomes, and perceived precision, trajectories were modeled as nonlinear smooth functions of time, allowed to differ across condition-specific series (and, for symptoms, across condition-by-assessment series), with random intercepts for participants and for participant-by-series terms (participant-by-assessment for symptoms) to capture within-person dependence. Residual autocorrelation was modeled using an AR(1) structure within each participant series, with restarts at gaps in the time index (Pinheiro et al., 1999). Heteroscedasticity was accommodated by allowing residual variance to differ across assessments (Pinheiro et al., 1999). Models were estimated in R Software (R Core Team, 2026) in the RStudio platform (Posit Team, 2025) using mgcv with REML/fREML estimation, and model adequacy was evaluated via gam.check and residual diagnostics (Wood, 2025). To support planned inference, complementary mixed-effects models were fit to obtain estimated marginal means at prespecified time points (baseline, midpoint, endpoint), simple time trends within assessment-by-condition strata, and multiplicity-adjusted pairwise contrasts (Lenth & Piaskowski, 2017).

## 4.1. Generalized Adaptive Mixed Modeling (GAMM)

### 4.1.1. *Hyperarousal Trajectories*

We analyzed longitudinal hyperarousal trajectories by applying *GAMM* to evaluate how daily hyperarousal events changed over time across three conditions (at-home control, physical intervention-only, digital + physical intervention). As shown in **Figure 5**, baseline-normalized mean MS values fluctuated over days, with the physical intervention-only group exhibiting larger late-study elevations and wider uncertainty bands than the at-home control and digital + physical intervention groups. To control for between-person differences in baseline hyperarousal frequency, we normalized each participant's daily values to a personal baseline defined as the mean of the first five observed (non-missing) days. We then computed a baseline-normalized rate ratio (ms_ratio = ms/baseline) and modeled the transformed outcome *log*(*ms_ratio* + 0.01) so that days with zero hyperarousal moments were retained in the analysis. Because repeated daily observations were nested within participants, the model included a participant-specific random intercept to account for within-person dependence (i.e., stable individual differences in overall hyperarousal level). An AR(1) residual structure was specified ($\rho = 0.30$) and restarted at gaps in the day index. To infer whether trajectories differed by condition, we compared (via an analysis of deviance) a model with a single shared nonlinear time function to one with condition-specific nonlinear time functions, testing whether allowing separate trajectories by condition improved model fit. The analysis included 465 person-days from 14 participants across 47 unique study days.

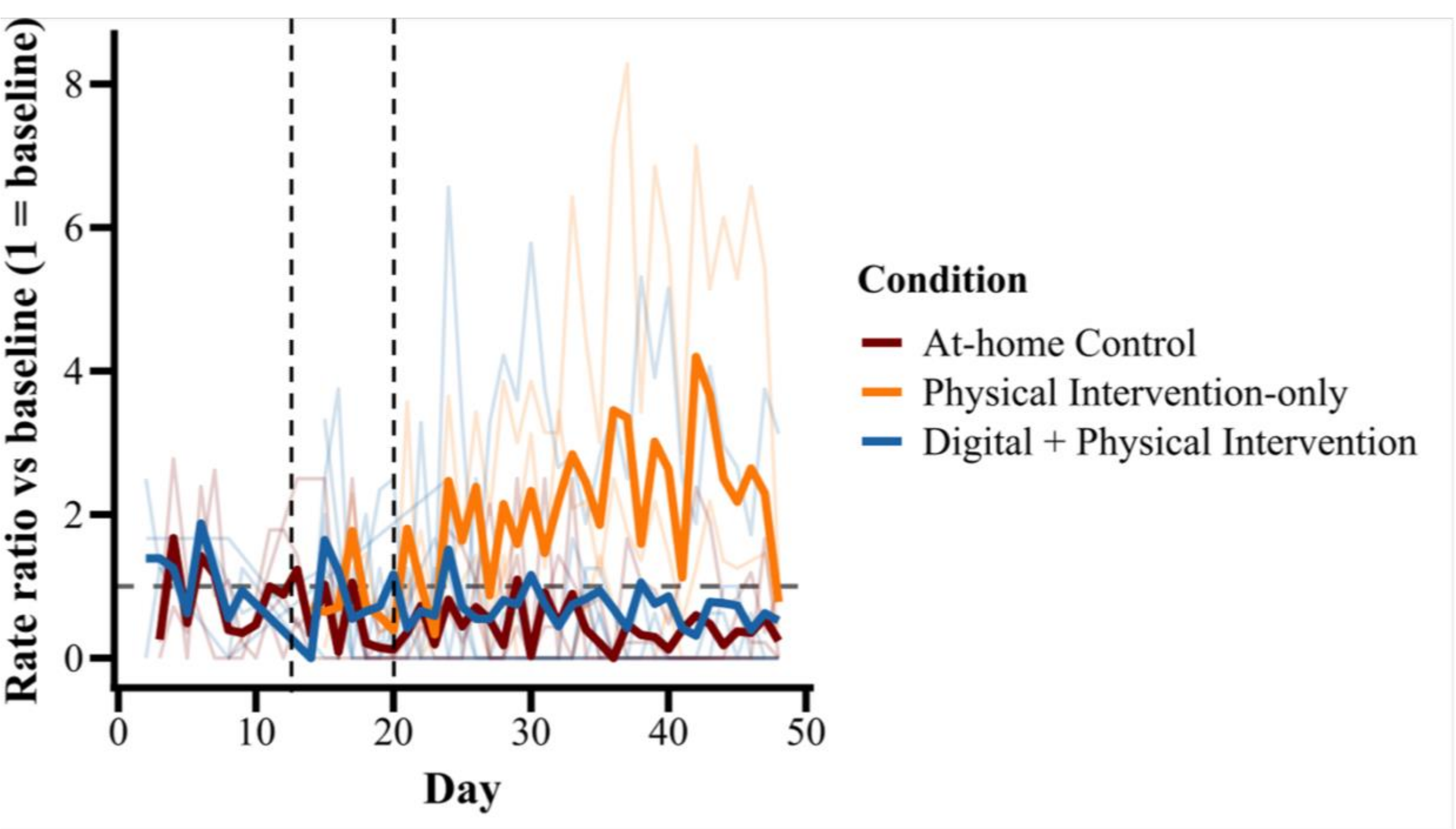

**Figure 5.** Baseline-normalized hyperarousal trajectories across days by condition. Daily values are expressed as a rate ratio relative to each participant's personal baseline (baseline = mean of the first five observed, non-missing days; horizontal dashed line at 1.0). Thick colored lines depict the condition-level daily means, while lighter individual traces show participant-specific day-to-day trajectories. Vertical dashed lines denote when the Texas Challenge occurred.

The overall model explained a moderate proportion of variance in the transformed outcome (adjusted $R^2 = .519$), accounted for 53.7% of the deviance, and provided evidence that time trajectories differed across conditions. A model with condition-specific nonlinear time curves fit significantly better than a model with a single common time curve across conditions, *ΔDeviance* = 57.00, *Δdf* = 3.74, $p$ = .002 (analysis of deviance), underscoring that the shape of change over days in normalized hyperarousal differed by condition. As seen in **Table 1**, on the transformed scale, the parametric contrasts for condition (relative to at-home control) were not statistically significant for both physical intervention-only ($p$ = .221) and digital + physical intervention ($p$ = .790). On the other hand, smooth time effects within condition showed a statistically reliable day-to-day pattern in the at-home control condition, $F(1.00, 1.00) = 7.07$, $p = .008$, and in the digital + physical intervention condition, $F(2.19, 2.74) = 8.30$, $p < .001$, whereas the physical intervention-only condition did not show a reliable nonlinear day effect, $F(1.72, 2.14) = 0.88$, $p = .387$. The participant random intercept was significant, indicating substantial between-person differences in overall hyperarousal levels, $F(10.29, 11.00) = 14.67$, $p < .001$; see (**Table 1**). **Figure 6** further illustrates this interpretation by presenting the GAMM-estimated trajectories with 95%CIs on the baseline-normalized scale. At-home control and digital + physical intervention conditions remain relatively low, exhibiting modest, structured temporal change, while the physical intervention-only condition follows a distinct trajectory, with a gradual increase in hyperarousal events later in the study.

**Table *1*:** GAMM Results for baseline-normalized hyperarousal moments.

| Model term | Estimate | SE | *t*-Test | *p*-Value |
|---|---|---|---|---|
| **Panel A. Fixed effects (parametric terms)** | | | | |
| Intercept (at-home control) | -2.499 | 0.857 | -2.92 | .004 |
| physical intervention-only vs at-home control | 1.622 | 1.322 | 1.23 | .221 |
| digital + physical intervention vs at-home control | -0.287 | 1.078 | -0.27 | .790 |
| **Panel B. Nonlinear time effects (smooth terms)** | | | | |
| **Smooth term** | **EDF** | **Ref. df** | ***F*-Test** | ***p*-Value** |
| Day trajectory: at-home control | 1.000 | 1.000 | 7.07 | .008 |
| Day trajectory: physical intervention-only | 1.716 | 2.139 | 0.88 | .387 |
| Day trajectory: digital + physical intervention | 2.192 | 2.744 | 8.30 | <.001 |
| Participant random intercept[a] | 10.289 | 11.000 | 14.67 | <.001 |

*Note.* Outcome was the log-transformed baseline-normalized hyperarousal rate ratio, $\log(ms_{\text{ratio}} + 0.01)$, where $ms_{\text{ratio}} = ms/\text{baseline}$ and baseline is the mean of each participant's first five observed (non-missing) days. **Panel A** reports parametric fixed-effect estimates; **Panel B** reports smooth terms for nonlinear change over day. EDF = estimated degrees of freedom (larger values indicate greater nonlinearity); reported $p$-values for smooth terms are approximate. Models were estimated via *f*REML and included a participant-specific random intercept; an AR(1) residual structure was specified and restarted at gaps in the day index (see text). [a]Implemented as a participant-specific random intercept in the GAMM.

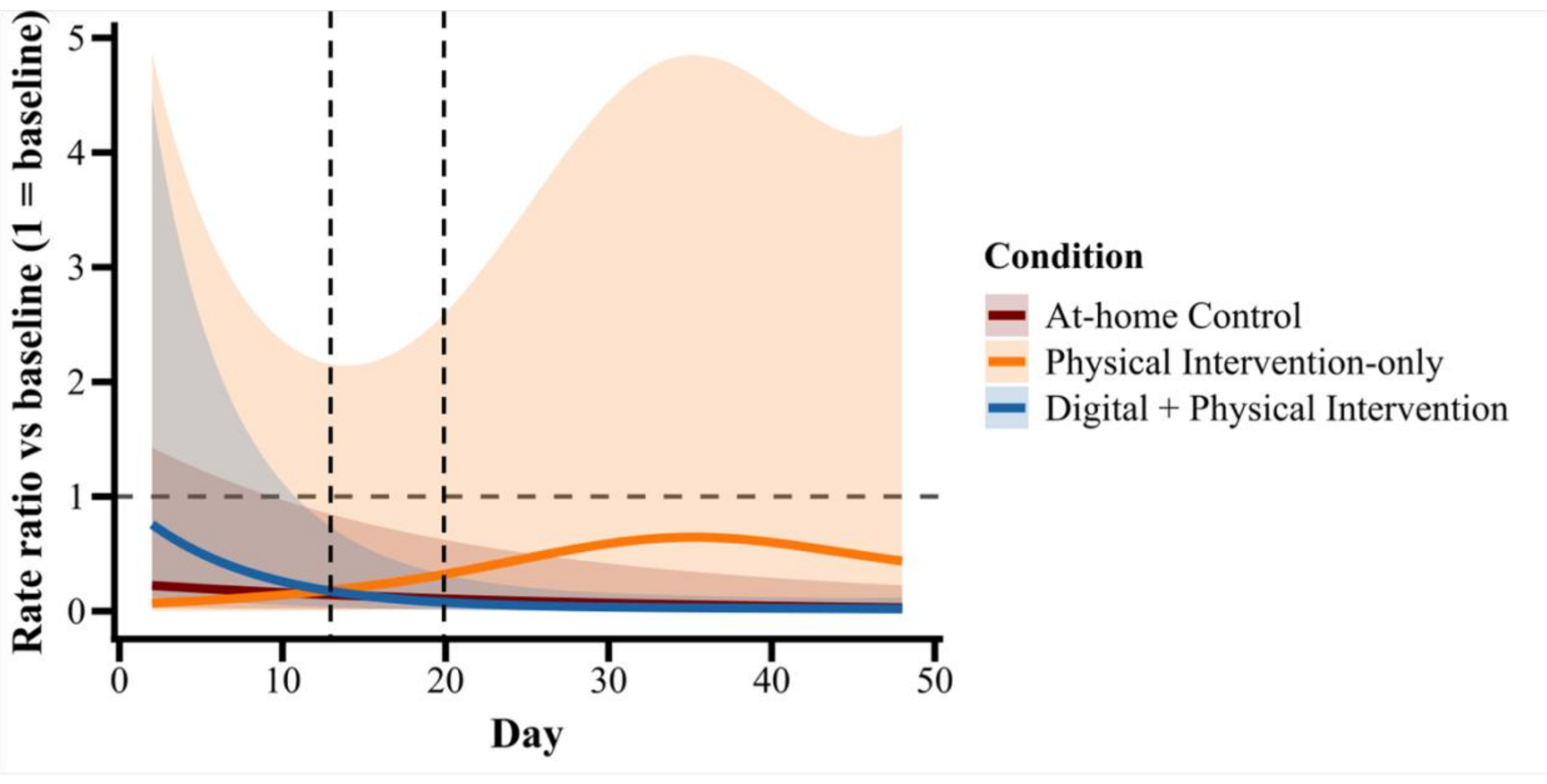

**Figure 6.** Model-estimated baseline-normalized hyperarousal trajectories by condition. Curves depict the GAMM-predicted daily MS rate ratio relative to each participant's personal baseline (baseline = mean of the first five observed, non-missing days; horizontal dashed reference line at 1.0). Shaded ribbons show 95% Confidence Intervals (CIs) around the population-level estimated trajectories for at-home control, physical intervention-only, and digital + physical intervention. Vertical dashed lines denote when the Texas Challenge occurred.

#### *4.1.2. Psychological Symptom Trajectories*

We analyzed longitudinal symptom trajectories using GAMMs to evaluate how three symptom scores, including GAD-7, PCL-5, and PHQ-8, changed over time as a function of the three conditions and the symptom assessments. From a descriptive statistics standpoint, **Figure 7** summarizes the observed group means over time, showing week-to-week trajectories for each assessment across the three conditions and providing an initial, model-free view of between-group separation and within-group variability prior to the GAMM-based estimates. The Texas Challenge occurred between weeks 2 and 3.

In the digital + physical intervention group, anxiety scores decreased from a baseline mean of 6.33 'Mild Anxiety') to 4.00 at Week 3 ('Minimal Anxiety'), reaching a temporary low at Week 4 (1.67, 'Minimal Anxiety') before rising slightly to 4.40 ('Minimal Anxiety') by Week 6. Thus, although fluctuations occurred after the event, mean anxiety levels remained within the minimal range at the end of the study. For depressive symptoms, the digital + physical intervention group decreased from 6.75 ('Mild Depression') at baseline to 2.25 ('Minimal Depression') during the ride period (Week 3), followed by a return to 5.60 (‘Mild Depression’) at Week 6. Although this represents a rebound relative to the mid-study low point, symptom levels remained below the commonly used clinical screening threshold of 10 throughout the study period. For PTSD symptoms, the digital + physical intervention group declined from 16.0 at baseline to 8.75 during the mid-study period, ending at 14.0 at Week 6. All values remained well below the commonly used screening threshold of 32 for probable PTSD.

The physical intervention-only group began with substantially higher baseline anxiety (21.0, ‘Severe Anxiety’), showed a marked reduction during the endurance challenge (Week 3 = 9.5, between Mild and Moderate Anxiety), and then increased greatly post-event (Week 4 = 21.0, ‘Severe Anxiety’), before decreasing again to 12.5 (‘Moderate Anxiety’) at Week 6. Despite post-event fluctuations, mean anxiety levels remained lower than baseline by the final assessment. As for depression symptoms, the physical intervention-only group again began with markedly elevated scores (24.0, ‘Severe Depression’) and showed a substantial reduction during the endurance challenge (Week 3 = 12.5, ‘Moderate Depression’), followed by increases at Week 4 (23.5, ‘Severe Depression’) and slight decline at Week 6 (15.0, ‘Moderately Severe Depression’). Despite post-event increases, mean depressive symptoms remained lower than baseline by the final assessment. With regard to PTSD symptoms, the physical intervention-only group began with very high baseline scores (80.0), decreased markedly during the endurance challenge (Week 3 = 39.0), and then increased again during the post-event period (Week 4 = 80.0; Week 6 = 62.67). Although scores remained above the threshold for probable PTSD throughout the study, the magnitude of the mid-study decrease represents the largest absolute change observed across the three groups.

The at-home control group showed a more gradual change in anxiety symptoms, declining from 11.0 (‘Moderate Anxiety’) at baseline to 7.0 (‘Mild Anxiety’) by Week 6. Similarly, the at-home control group showed gradual reductions in depressive symptoms across the study period, declining from 11.0 (‘Moderate Depression’) to 6.67 (‘Mild Depression’) by Week 6. Finally, the at-home control group demonstrated gradual improvement in PTSD symptoms, declining from 32.0 at baseline (consistent with the commonly used screening threshold for probable PTSD) to 17.33 by Week 6, falling below the screening threshold by the final assessment.

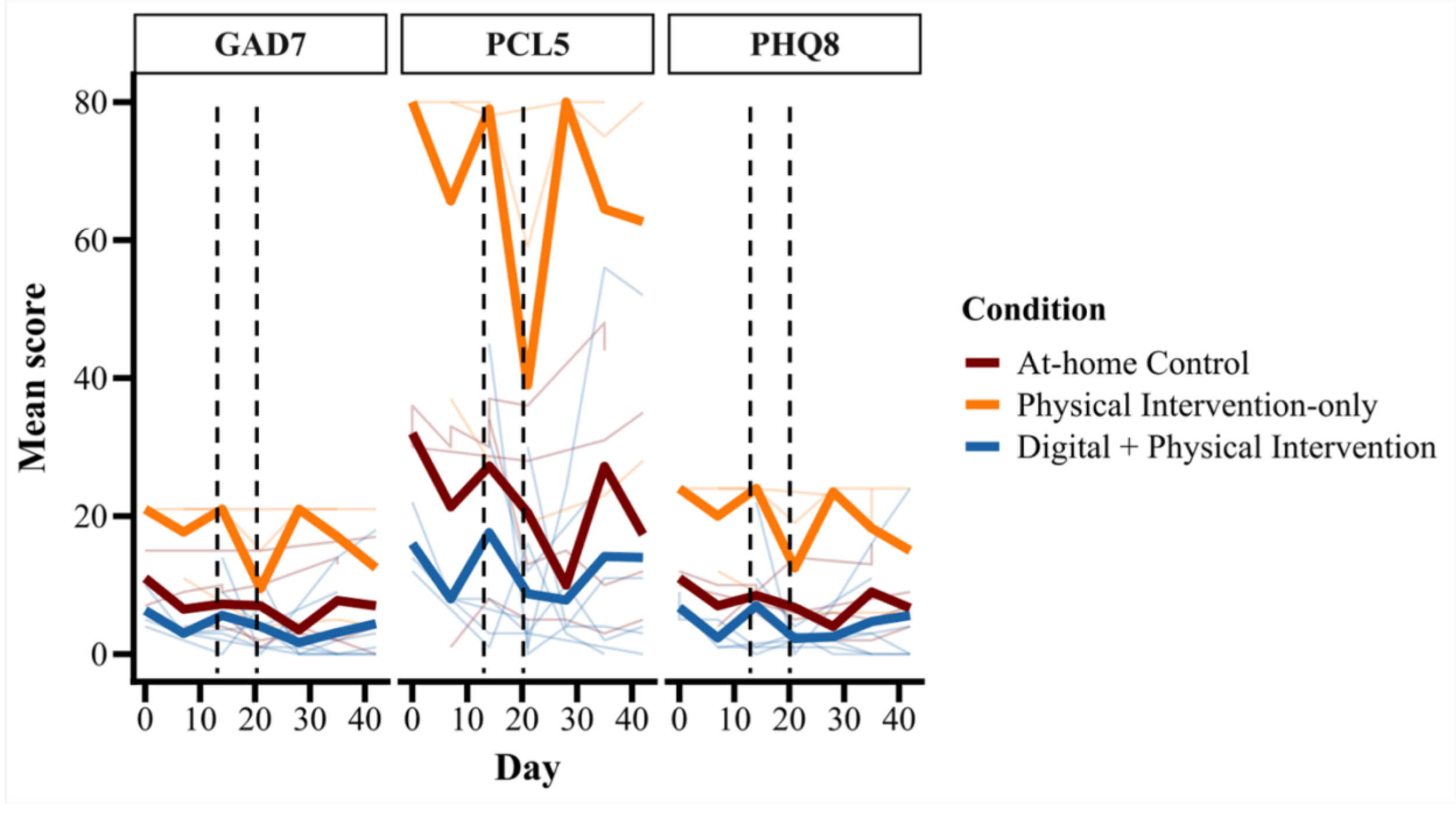

**Figure 7.** Observed self-assessment trajectories over time by condition and assessment. Panels show weekly GAD-7, PCL-5, and PHQ-8 scores for each condition. Thin lines depict individual participant trajectories, and thick solid lines show the corresponding weekly condition means. Vertical dashed lines denote when the Texas Challenge occurred.

To allow for assessment-specific dispersion, residual variance was permitted to differ across assessments (i.e., separate residual standard deviations for GAD-7, PCL-5, and PHQ-8). We also specified an AR(1) residual correlation within each participant-by-assessment series; the estimated autocorrelation was approximately zero, indicating negligible residual serial dependence after accounting for fixed and smooth effects. Models were estimated using *f*REML. For planned inferential follow-ups, we fit a companion linear mixed-effects model with the same random-effects structure to test time-by-condition-by-assessment effects and to obtain simple time slopes and estimated marginal means at prespecified time points (baseline, midpoint, endpoint), with Tukey-adjusted pairwise contrasts within each assessment and time point.

Overall, the GAMM explained 55.2% of the variance in symptom scores (Adjusted $R^2 = 0.552$, n = 211), and the findings are summarized in **Table 2**. Nonlinear time effects were detected only in the digital + physical intervention condition for GAD-7, $F(2.02,2.02) = 3.38$, $p = 0.038$, and PHQ-8, $F(2.35,2.35) = 3.29$, $p = 0.036$, whereas all other condition-by-assessment smooths were not statistically reliable (all $p \geq 0.287$). In comparisons at baseline, mid, and end, PCL-5 scores were consistently higher in the physical intervention-only condition than in the at-home control and digital + physical intervention at each timepoint (Tukey-adjusted *ps* ≤ 0.0039), whereas no condition differences were observed for GAD-7 or PHQ-8 at any timepoint (Tukey-adjusted *ps* ≥ 0.11). A linear mixed-effects follow-up test of the condition by assessment by time interaction was not significant, LR=1.09, $p = 0.896$, suggesting that differential change was better characterized by nonlinear trajectory features captured by the GAMM smooths rather than by differences in linear slopes. Overall, these findings are illustrated in **Figure 8**, which shows the GAMM-predicted self-assessment trajectories by condition and assessment, showing the fitted mean curves for GAD-7, PCL-5, and PHQ-8 across study weeks with 95% CIs, which is complementary to the descriptive group means in **Figure 7** by depicting the estimated nonlinear time patterns implied by the GAMM.

**Table 2.** GAMM Results for symptom scores across assessments.

| Model term | Estimate | SE | t-Test | p-Value |
|---|---|---|---|---|
| **Panel A. Fixed effects (parametric terms)** | | | | |
| Intercept (physical intervention-only, GAD-7) | 15.668 | 5.173 | 3.029 | .0028 |
| at-home control | -7.839 | 6.841 | -1.146 | .2533 |
| digital + physical intervention | -11.207 | 6.184 | -1.812 | .0715 |
| PCL-5 | 45.801 | 4.815 | 9.511 | < .001 |
| PHQ-8 | 2.259 | 4.202 | 0.537 | .5916 |
| at-home control × PCL-5 | -31.888 | 6.356 | -5.017 | < .001 |
| digital + physical intervention × PCL-5 | -36.639 | 5.750 | -6.372 | < .001 |
| at-home control × PHQ-8 | -2.626 | 5.563 | -0.472 | .6374 |
| digital + physical intervention × PHQ-8 | -1.912 | 5.022 | -0.381 | .7039 |
| **Panel B. Nonlinear time effects (smooth terms)** | | | | |
| **Smooth term** | **EDF** | **Ref. df** | **F-Test** | **p-Value** |
| $s(d_c)$: physical intervention-only.GAD-7 | 1.000 | 1.000 | 0.620 | .4319 |
| $s(d_c)$: at-home control.GAD-7 | 1.000 | 1.000 | 0.144 | .7049 |
| $s(d_c)$: digital + physical intervention.GAD-7 | 2.015 | 2.015 | 3.375 | .0375 |
| $s(d_c)$: physical intervention-only.PCL-5 | 1.000 | 1.000 | 0.847 | .3585 |
| $s(d_c)$: at-home control.PCL-5 | 1.000 | 1.000 | 0.041 | .8406 |
| $s(d_c)$: digital + physical intervention.PCL-5 | 1.788 | 1.787 | 1.740 | .2873 |
| $s(d_c)$: physical intervention-only.PHQ-8 | 1.000 | 1.000 | 0.866 | .3532 |
| $s(d_c)$: at-home control.PHQ-8 | 1.000 | 1.000 | 0.084 | .7716 |
| $s(d_c)$: digital + physical intervention.PHQ-8 | 2.345 | 2.345 | 3.286 | .0357 |

***Note.*** Outcome was symptom score ($s$) measured weekly across assessments (GAD-7, PCL-5, PHQ-8). **Panel A** reports parametric fixed-effect estimates; **Panel B** reports smooth terms for nonlinear change over week within each condition ×assessment series. EDF = estimated degrees of freedom (larger values indicate greater nonlinearity); reported $p$-Values for smooth terms are approximate. Models were estimated via $f$REML and included random intercepts for participants and for the participant-by-assessment series; residual variance was permitted to differ across assessments. An AR(1) residual structure was specified within each segmented participant-by-assessment series; the estimated autocorrelation was approximately zero.

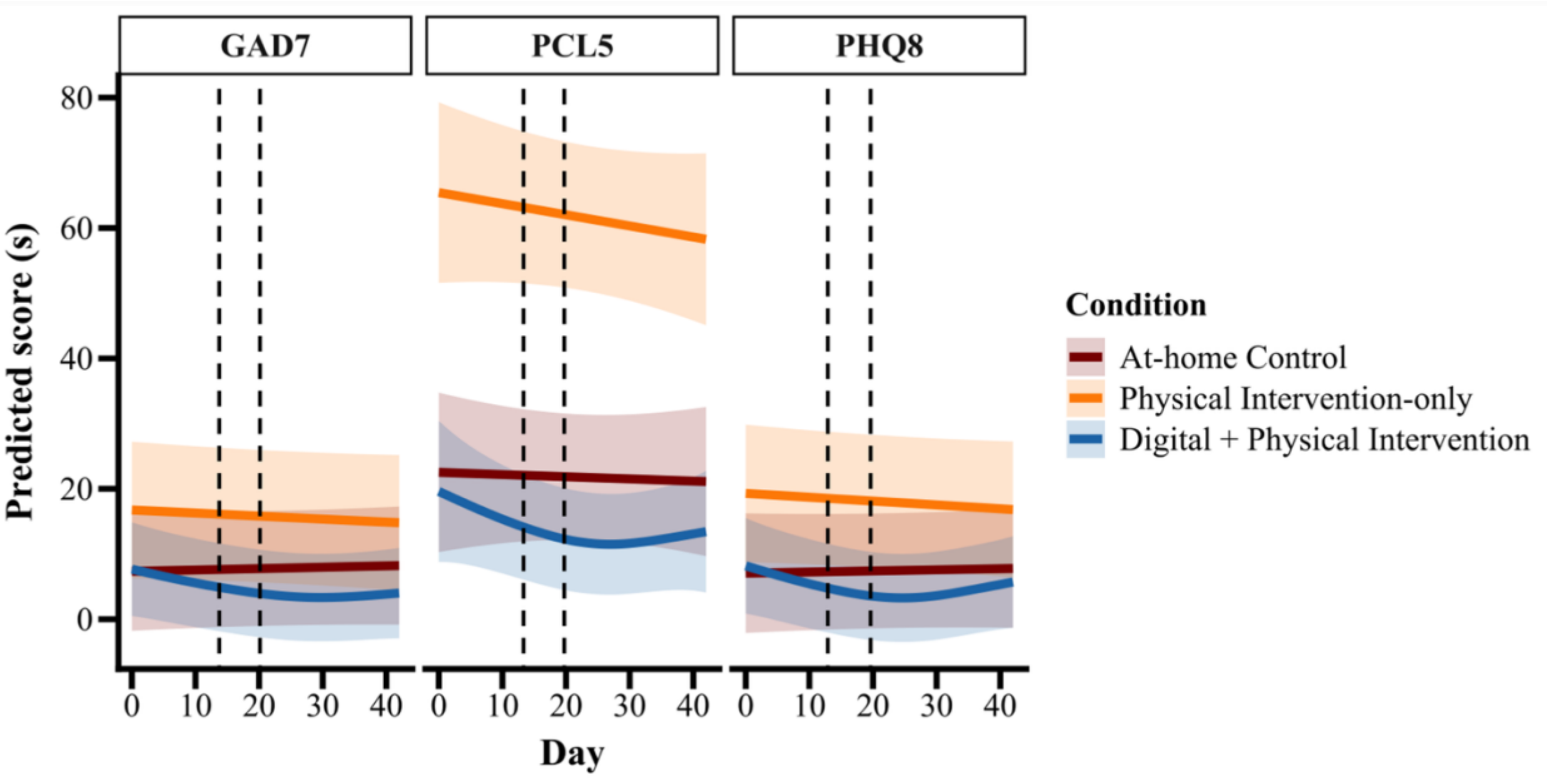

**Figure 8**. GAMM-predicted self-assessment trajectories over time by condition and assessment. Panels show model-based predicted scores for GAD-7, PCL-5, and PHQ-8 across study weeks for each of the three conditions. Solid lines represent the fitted mean trajectories from the GAMM; shaded bands indicate 95% CIs around the predicted means. Vertical dashed lines denote when the Texas Challenge occurred.

#### ***4.1.3. Perceived Precision***

Across the full study period of up to 48 days, a total of 3,178 hyperarousal-related events were recorded, including system- and ML-detected events and participant self-reports. Of these events, 1,519 were confirmed detections (“Yes”), 907 were rejected detections (“No”), and 752 were self-reported events. On average, participants generated 227.0 events (SD = 246.87), with an average of 92.62 confirmed detections (SD = 170.30), 82.07 rejected detections (SD = 80.77), and 55.86 self-reported events (SD = 115.56). **Figure 9** illustrates the distribution of response types across participants. Substantial between-participant variability was observed in overall event counts and response patterns.

Self-reported events occurred less frequently than system-detected events across participants, suggesting that most hyperarousal episodes were captured through automated detection rather than participant-initiated reporting. Based on the analysis, the mean perceived precision across all participants was 43.08% (SD = 36.31%), with a median of 39.15% and a range of 1.0% to 100.0%. **Figure 10** shows the perceived precision for each participant from lowest to highest.

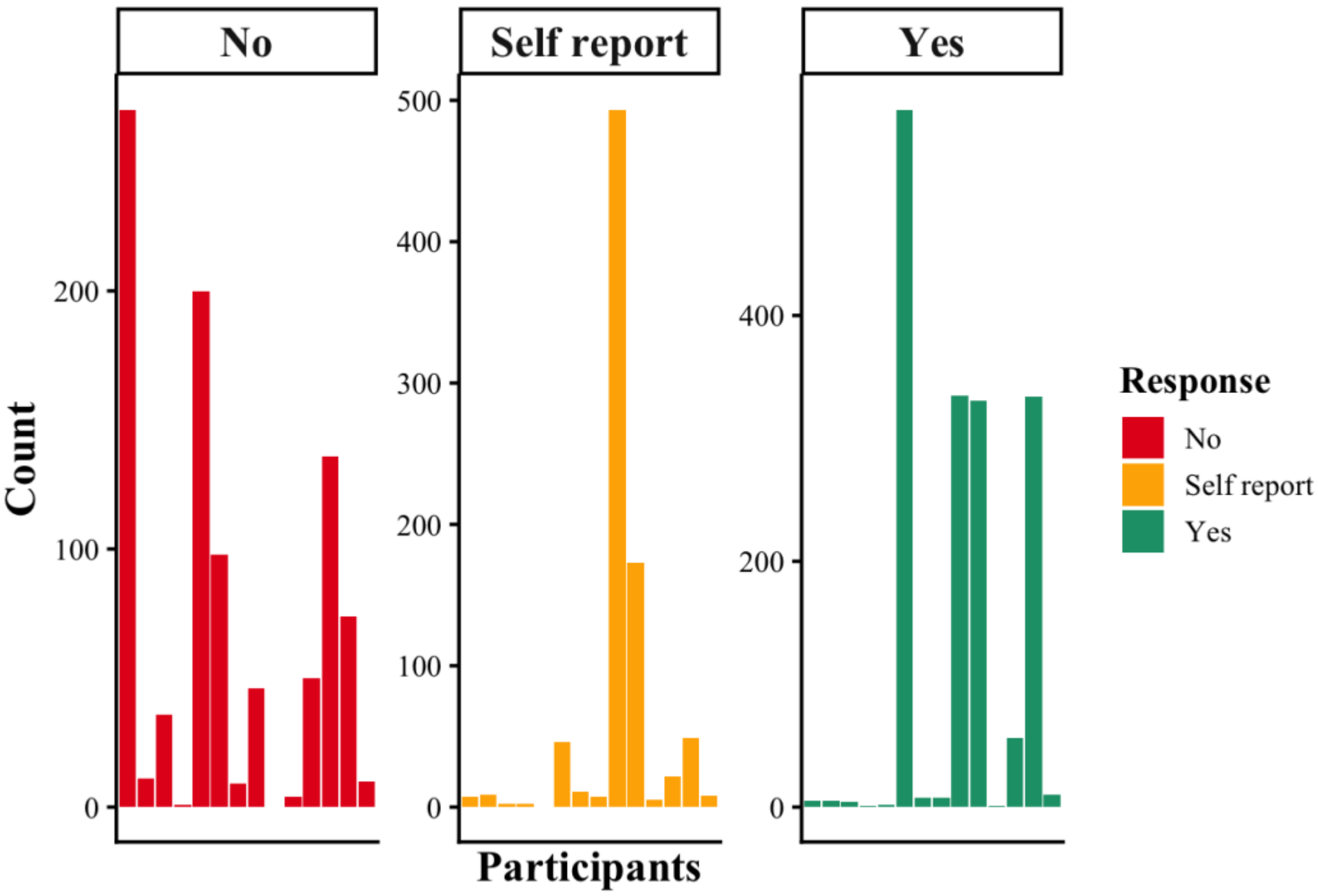

**Figure 9:** Count of “Yes”, “No”, and self-reported events for each participant.

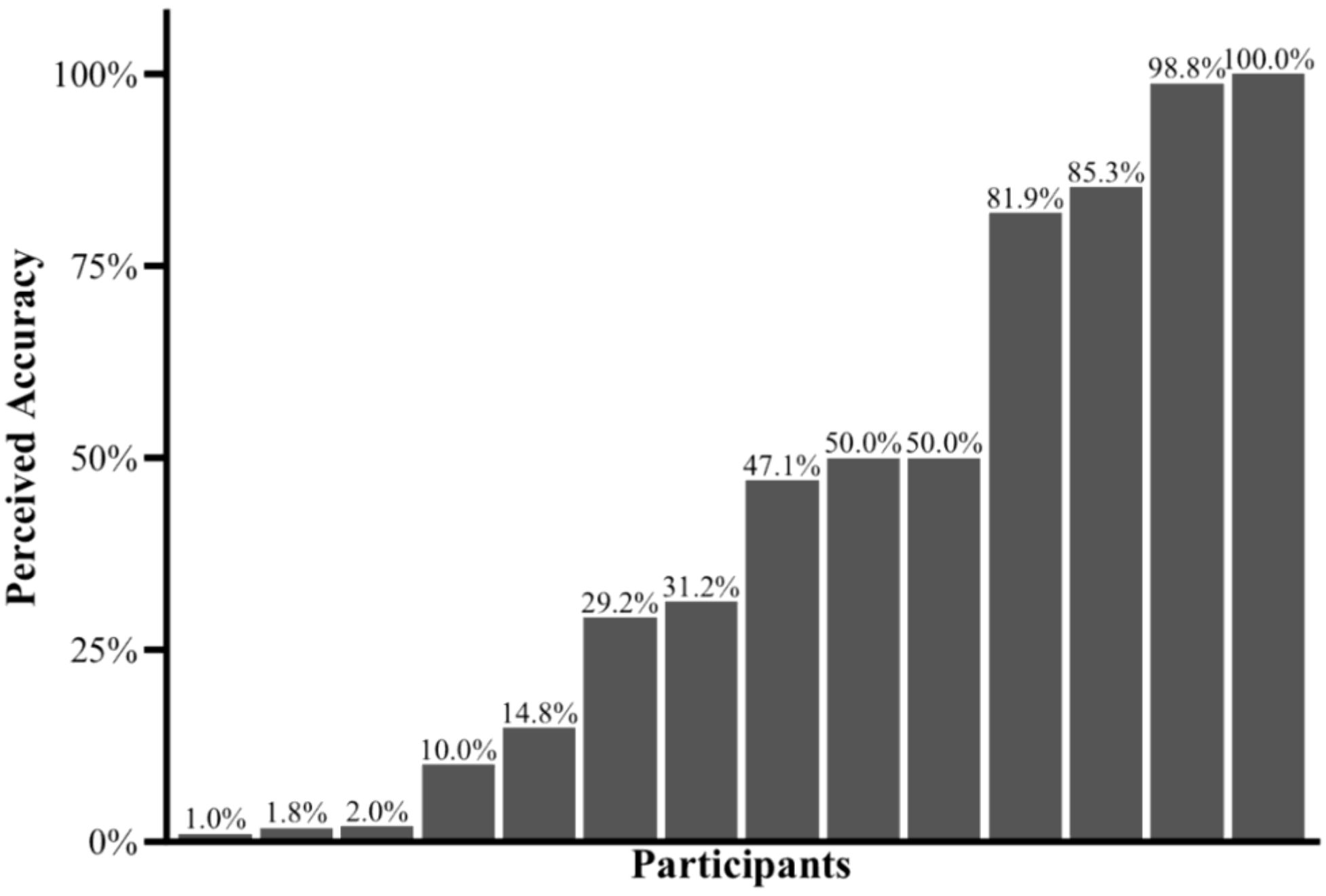

**Figure 10:** Overall Mean Perceived Accuracy for each Participant

To investigate how perceived precision changed over time and to assess whether these trajectories varied across experimental conditions, we applied GAMMs to the daily counts of confirmed and rejected detections for each participant (i.e., perceived precision). For each participant on each study day, we modeled the number of confirmed detections (“Yes”) and rejected detections (“No”) as binomial outcomes, specifying a random intercept for each participant to account for stable individual differences in response patterns. Time was modeled as a smooth function of study day to capture potential nonlinear trends. To evaluate whether the pattern of change differed by condition, we compared two models (**Table 3**): one in which the smooth function of time was shared across all conditions, and another in which the time smooth was estimated separately for each condition. The condition-specific model provided a significantly better fit than the shared-smooth model, $p$ = .007; henceforth, we retained the condition-specific model for inference and thereby theresults from the binomial GAMM with a logit link to model perceived precision over time, with participant included as a random effect and condition entered as a fixed effect. **Figure 11** visualizes the substantial day-to-day variability in participants' perceived precision, along with condition-level trajectories over time which indicates that intervention group's trajectory stabilizes later duration of the task.

**Table 3.** Model comparison for perceived precision trajectories across conditions (binomial GAMM).

| Model | Resid. df | Resid. deviance | Δdf | Δdeviance | p-Value |
|---|---|---|---|---|---|
| Shared time smooth | 303.44 | 379.71 | — | — | — |
| Condition-specific smooths | 297.18 | 361.70 | 6.2533 | 18.01 | 0.007384 |

Note. Both models were fit with a binomial family (logit link) to participant-day confirmation counts: cbind(detected yes, fail). Both included a random intercept for participants, and the condition-specific model additionally allowed time to vary by condition, specified as the number of days, *with a p-value from an* analysis of deviance ($\chi^2$ test).

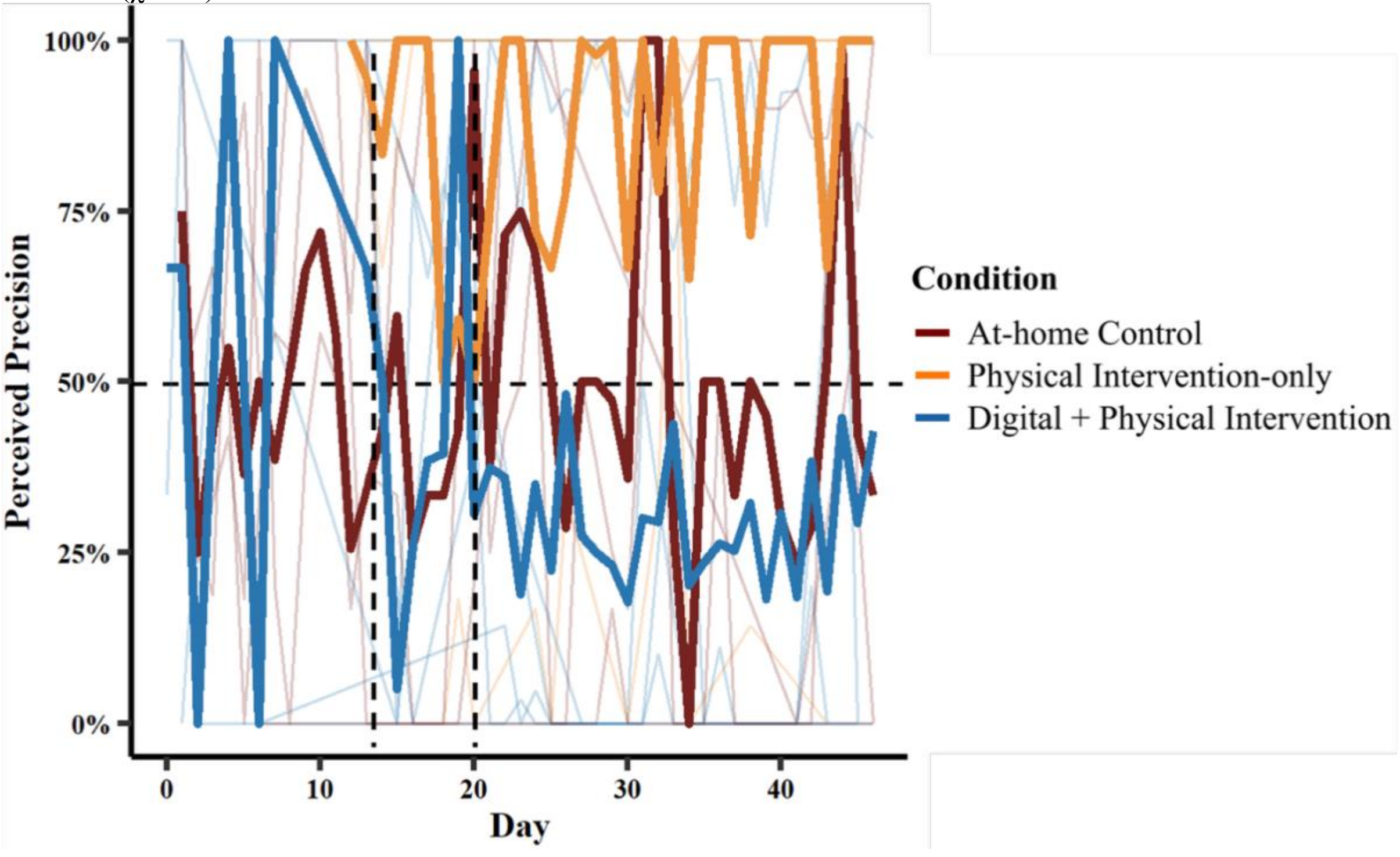

**Figure 11:** Individual and condition-level trajectories of perceived precision over time. Thin lines represent each participant's daily perceived precision, and thick lines represent the condition-level daily mean trajectory. The dashed horizontal line indicates the reference precision level (50%). Vertical dashed lines denote when the Texas Challenge occurred.

As seen in **Table 4**, the parametric condition contrasts (physical intervention-only and digital + physical intervention relative to at-home control, reference group) were not statistically significant (*ps* > .05). In contrast, the time-varying smooth terms indicated that the trajectory of perceived precision over time was significant for the at-home control condition ($p < .001$), not significant for physical intervention-only ($p > .05$), and marginal for digital + physical intervention ($p = .061$). The participant-level random-effect smooth was also significant ($p < .001$), consistent with substantial between-participant variability.

These results indicate a clear time trend in at-home control, no systematic change in physical intervention-only, and a borderline time trend in digital + physical intervention, while average condition differences were not statistically significant ($p > .05$). **Figure 12** visualizes these model-estimated trajectories, showing a clear temporal pattern in at-home control, little systematic change in physical intervention-only, and a weaker/borderline temporal pattern in Intervention, despite no evidence of overall mean differences between conditions.

**Table 4.** Binomial GAMM results with condition-specific time smooths (perceived precision).

| Model term | Estimate | SE | z-Test | p-Value |
|---|---|---|---|---|
| **Panel A. Fixed effects (parametric terms)** | | | | |
| Intercept (at-home control) | -0.4982 | 1.5311 | -0.325 | 0.745 |
| physical intervention-only | 3.5004 | 2.3836 | 1.469 | 0.142 |
| digital + physical intervention | -1.0311 | 1.9279 | -0.535 | 0.593 |
| **Panel B. Nonlinear time effects (smooth terms)** | | | | |
| **Smooth term** | **EDF** | **Ref. df** | **$\chi^2$ -Test** | **p-Value** |
| *s*(DayNum): at-home control | 2.790 | 3.463 | 29.701 | <.001 |
| *s*(DayNum): physical intervention-only | 1.065 | 1.127 | 3.028 | 0.1047 |
| *s*(DayNum): digital + physical intervention | 4.226 | 5.138 | 10.771 | 0.0609 |
| *s*(ParticipantID): RE | 10.426 | 11.000 | 558.181 | <.001 |

Note. Binomial GAMM with logit link fit to cbind(detected yes, fail). Condition-specific trajectories modeled via s(DayNum, by=condition) with participant random intercepts. n = 323 participant-days. Adjusted R2 = 0.879; deviance explained = 84.8%; REML = 329.67.

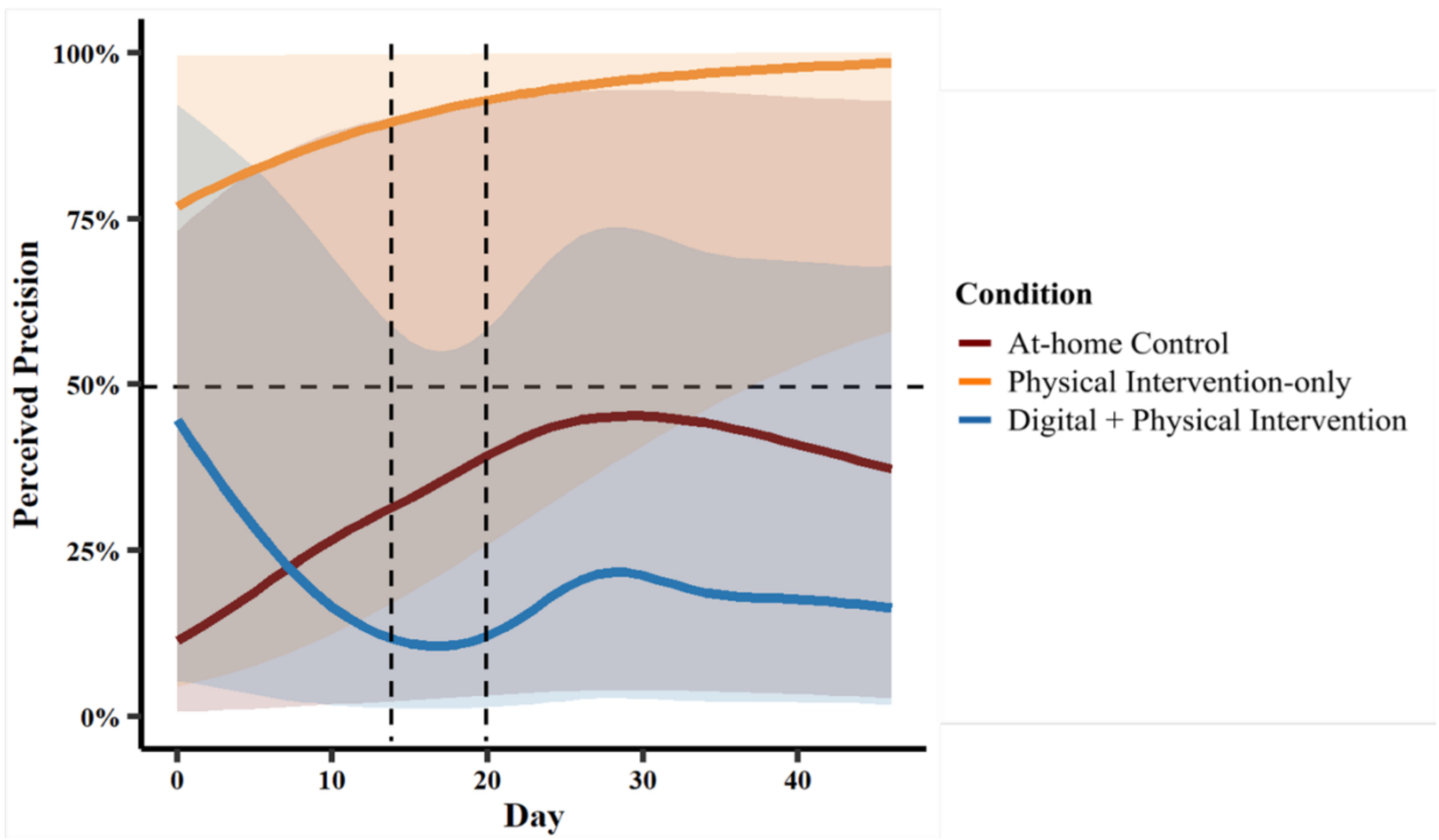

**Figure 12.** Condition-specific trajectories of perceived precision over time. Lines show population-average predicted perceived precision from the binomial GAMM (condition-specific smooths of Day; participant included as a random effect), with shaded bands represents 95% CI. The dashed horizontal line marks the reference precision level (50%). Vertical dashed lines denote when the Texas Challenge occurred.

To further examine whether perceived precision varied as a function of baseline symptom severity, we aggregated perceived precision to weekly mean values and paired these with weekly symptom assessments (PCL-5, GAD-7, and PHQ-8). Participants were stratified into low- and high-severity groups for each assessment (low severity was defined as PCL-5 < 31; GAD-7 < 10; PHQ-8 < 10), and binomial GAMMs were fit to model weekly perceived precision trajectories as a function of severity and time. The high severity group included four participants, while the low severity group included 10 participants. **Figure 13** presents the weekly daily perceived precision trajectories by severity group, showing substantial between-participant variability but a consistent pattern in which high-severity participants tended to exhibit higher perceived precision over time.

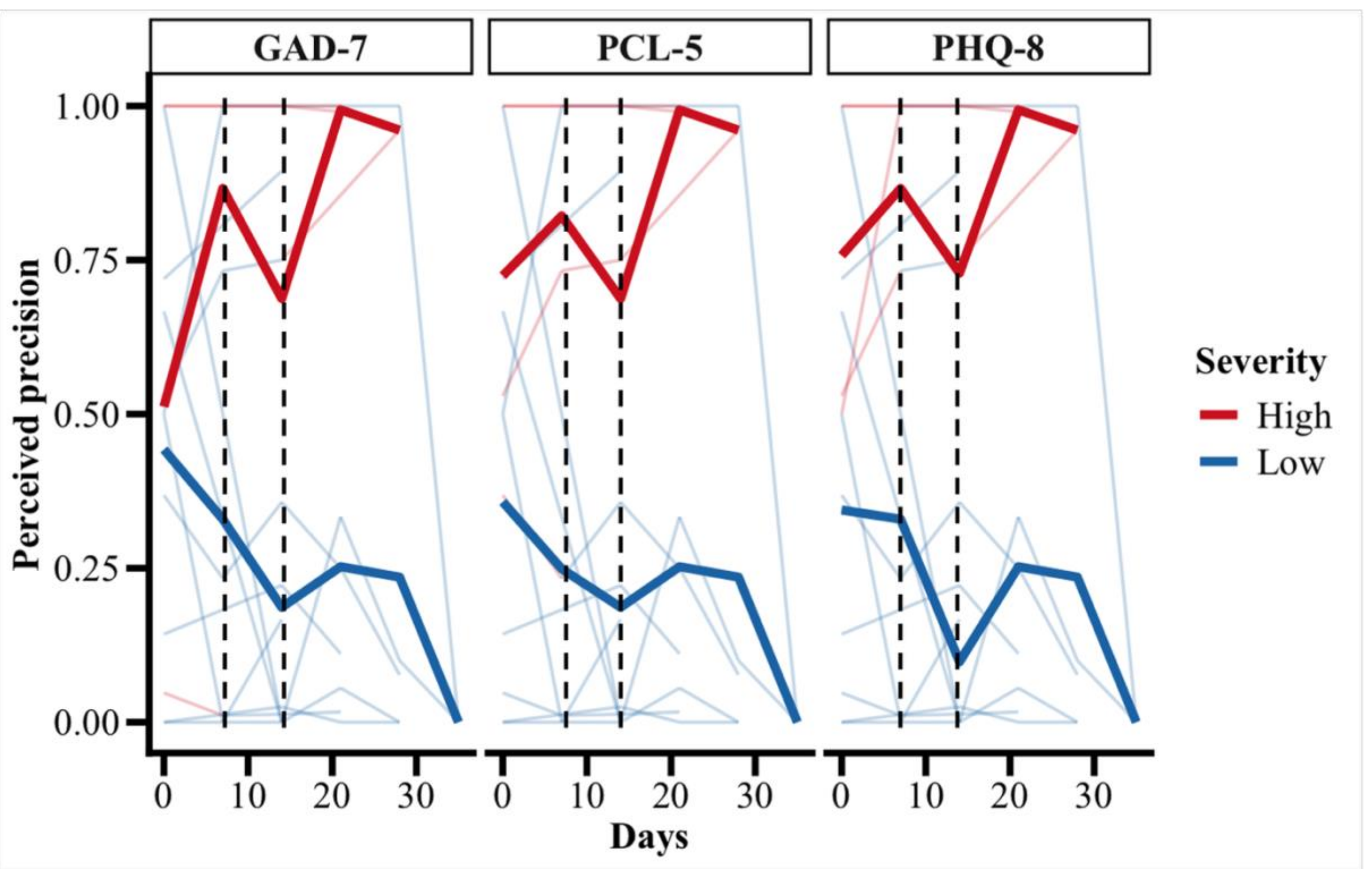

**Figure 13:** High and low-level severity trajectories of perceived precision over time. Thin lines represent each participant's daily perceived precision, and thick lines represent the condition-level daily mean trajectory. Vertical dashed lines denote when the Texas Challenge occurred.

Across all three assessments, higher symptom severity was associated with greater weekly perceived precision. As summarized in **Table 5** (Panel A), participants in the high-severity groups exhibited significantly higher perceived precision than those in the low-severity groups for PCL-5 ($p < .001$), GAD-7 ($p < .001$), and PHQ-8 ($p < .001$). These results indicate that individuals reporting higher weekly symptom burden were more likely to confirm system-detected hyperarousal events.

Nonlinear time effects in weekly perceived precision were also observed (**Table 5**; Panel B). For PCL-5, significant smooth terms were detected in both severity groups (low: $\chi^2 = 10.87$, $p = .009$; high: $\chi^2 = 7.58$, $p = .016$), indicating changing weekly trajectories in perceived precision. For GAD-7 and PHQ-8, the high-severity groups showed strong increasing trends over time (both $p < .001$), whereas the low-severity groups exhibited weaker but still significant nonlinear patterns ($p = 0.024$ and $p = 0.050$, respectively). **Figure 14** presents the GAMM-predicted weekly trajectories with 95% confidence intervals, illustrating a consistent upward trend in perceived precision among high-severity participants compared to relatively stable patterns in low-severity groups.

**Table 5:** GAMM Results for perceived precision by symptom severity.

| Model term | Estimate | SE | z-Test | p-Value |
|---|---|---|---|---|
| **Panel A. Fixed effects (parametric terms)** | | | | |
| Intercept (Low severity, PCL-5) | -0.74 | 0.809 | -0.915 | 0.36 |
| High severity x PCL-5 | 1.268 | 0.339 | 3.74 | < .001 |
| Intercept (Low severity, GAD-7) | -0.975 | 0.737 | -1.324 | 0.186 |
| High severity x GAD-7 | 2.086 | 0.48 | 4.347 | < .001 |
| Intercept (Low severity, PHQ-8) | -0.727 | 0.778 | -0.935 | 0.35 |
| High severity x PHQ-8 | 1.066 | 0.289 | 3.688 | < .001 |
| **Panel B. Nonlinear time effects (smooth terms)** | | | | |
| **Smooth term** | **EDF** | **Ref. df** | **$\chi^2$-Test** | **p-Value** |
| s(day): Low.PCL-5 | 2.278 | 2.609 | 10.871 | 0.009 |
| s(day): High.PCL-5 | 1.689 | 1.968 | 7.578 | 0.016 |
| s(day): Low.GAD-7 | 2.25 | 2.59 | 8.292 | 0.024 |
| s(day): High.GAD-7 | 1.00 | 1.00 | 22.194 | < .001 |
| s(day): Low.PHQ-8 | 1.64 | 1.979 | 5.634 | 0.050 |
| s(day): High.PHQ-8 | 1.00 | 1.00 | 26.882 | < .001 |

***Note.*** Outcome was symptom score ($s$) measured weekly across assessments (GAD-7, PCL-5, PHQ-8). **Panel A** reports parametric fixed-effect estimates; **Panel B** reports smooth terms for nonlinear change over week within each condition ×assessment series. EDF = estimated degrees of freedom (larger values indicate greater nonlinearity); reported $p$-Values for smooth terms are approximate. Models random intercepts for participants and for the participant-by-assessment series; residual variance was permitted to differ across assessments. An AR(1) residual structure was specified within each segmented participant-by-assessment series; the estimated autocorrelation was approximately zero.

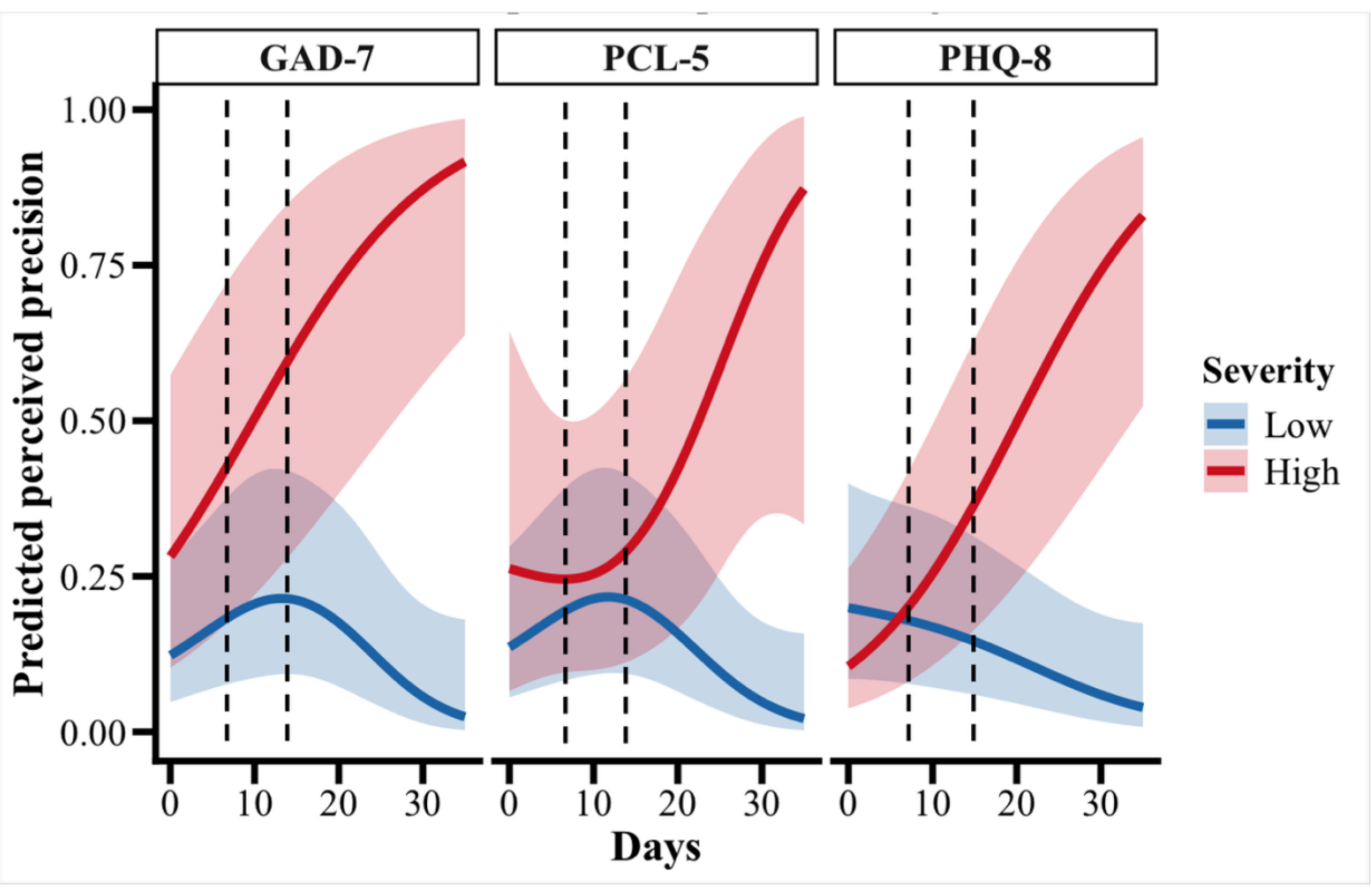

**Figure 14:** Severity-specific trajectories of perceived precision over time. Lines show population-average predicted perceived precision from the binomial GAMM (severity-specific smooths of day; participant included as a random effect), with shaded bands representing 95% CIs. Vertical dashed lines denote when the Texas Challenge occurred.

### 4.2. Veteran Perspectives and Takeaways

Qualitative feedback regarding the smartwatch detection tool was collected for participants across all conditions. Feedback highlighted themes related to the overall experience, perceived precision, post-detection expectations, and adoption barriers (**Table 6**); the results below present participants' perspectives through representative quotations. Participants largely described the detection tool as beneficial and easy to follow in daily life, most often because it increased awareness and mindfulness of hyperarousal/high-stress moments. A smaller subset reported that reminders were not helpful.

Perceived precision was generally favorable: multiple participants reported that detection was accurate and expressed trust in the detection capability. At the same time, some participants described the tool as overly sensitive and questioned accuracy, noting that alerts could occur during moments they did not perceive as stressful. Perceptions of false alarms varied considerably across participants, suggesting meaningful between-user differences in how model outputs aligned with lived experience and contextual factors.

Expectations about what should happen after detection varied. Some participants expected or desired immediate, actionable coping support delivered at the moment of detection (e.g., rapid access to breathing or grounding techniques), whereas others emphasized that awareness alone was sufficient. Another subset preferred minimal interaction, placing greater emphasis on alert modality (e.g., vibration or tone) and simple recording/logging features. A few participants indicated that the current implementation already aligned with their expectations, while others were uncertain about what post-detection support should entail.

Across responses, participants tended to prefer notifications that were brief and easy to interpret. However, desired content ranged from coaching-style guidance to minimal prompts intended primarily to facilitate self-reflection and situational awareness.

Finally, participants indicated willingness to use the tool daily, typically contingent on usability improvements. Battery drain and the need for frequent charging emerged as a prominent barrier, followed by concerns related to technical reliability (e.g., glitches or friction associated with charging, uploading, or app performance). In addition, responses suggested that interacting with notifications could be context-dependent, with professional settings sometimes making in-the-moment attention to alerts and response actions more difficult.

**Table 6:** Key Qualitative Findings Regarding Detection

| Factor | Key Findings | n | N | % | Representative Participant Quote |
|---|---|---|---|---|---|
| **Overall Experience** | Positive overall experience | 10 | 12 | 83.3 | "It was a great experience completing the daily task, which helped me stay on track." – P6. |
| | The tool increased awareness/ mindfulness | 10 | 12 | 83.3 | "It was good to have. I was able to see when I was having stressful moments throughout the day. The app was keeping me mindful of where I was at with my stress levels throughout my day and be able to use my coping mechanisms. It made me very mindful |
| | Reminders not helpful | 2 | 12 | 16.7 | |
| | | | | | of how I was doing throughout the day." – P12. |
| **Perceived Precision** | Accurately detects hyperarousal/ high-stress | 8 | 12 | 66.7 | "Yes it was accurate 80% of the time" – P7.<br>"Yes, a couple times it hit before I was even aware what was going on in my body" – P 9. |
| | Trust in detection capability | 7 | 12 | 58.3 | "Sometimes 6 or more [false alarms]. Generally I would say more than half were false alarms." – P13. |
| | Questioned accuracy / overly sensitive | 5 | 12 | 41.7 | |
| **Post-Detection Expectation** | Actionable coping support expected | 5 | 12 | 41.7 | "Maybe a coach type voice that guide you through stressful moment" – P1 |
| | Awareness-only sufficient | 2 | 12 | 16.7 | "I never expect anything except the alert and then do self-reflection" – P9 |
| | Minimal interaction/modality focus | 2 | 12 | 16.7 | "The alerts that I received were good to remind me of where I was at." – P12 |
| | Current implementation matched expectations. | 1 | 12 | 8.3 | "I wear hearing aids and perhaps the ability to Bluetooth a brief calming sound to them would be beneficial. It might also help redirect me from the stressful event." – P13. |
| | Uncertain about desired support | 2 | 12 | 16.7 | |
| **Acceptance & Barriers** | Willing to use | 6 | 12 | 50 | "I would use but it took a lot of battery and I would have to charge my watch during the day." – P6. |
| | Battery drain / frequent charging barrier | 4 | 12 | 33.3 | "Yes, recharging, uploading files and after watching videos the app would get a bit glitchy" – P7. |
| | Technical reliability issues barrier | 2 | 12 | 16.7 | |

Participants in the digital intervention group who completed the exit survey (N=5) also provided feedback on the educational content and engagement tools embedded within the FWD application, including videos, breathing exercises, music, and journaling features (**Table 7**). Overall, participants expressed generally positive perceptions of the educational content, with several describing it as helpful, informative, and well-suited for individuals earlier in their stress-management journey. Some participants emphasized that the educational materials would be particularly valuable for veterans and first responders who had not previously received formal training or support related to stress and mental health.

At the same time, perceptions of digital intervention content varied depending on participants' prior experience with stress management. A subset of participants reported that the educational content did not provide new information for them personally, as they had already developed their own coping strategies over time. This view could be a function of age, as many participants in the study were above 55 years old. Additionally, these participants are all established in the Project Hero community, which suggests they are resource seeking and actively manage their conditions. These participants nevertheless acknowledged that the content could be beneficial for others with less experience managing stress.

Feedback on engagement tools was similarly mixed but generally favorable. Breathing exercises, music, and video content were most frequently mentioned as useful, particularly during moments of elevated stress. Participants who engaged with these features highlighted their role in promoting relaxation and situational awareness. In contrast, journaling tools were used less frequently, and some participants reported limited recall or engagement with this feature. Some participants suggested potential improvements to enhance usability, including the addition of audio-guided breathing exercises and expanded multimedia content.

**Table 7**. Key Qualitative Findings Regarding Digital Intervention Content

| Factor | Key Findings | n | N | % | Representative Participant Quote |
|---|---|---|---|---|---|
| **Educational Content** | Positive perception of educational content | 4 | 5 | 80 | "I really like the educational content." – P2 |
| | Helpful for individuals newer to stress management | 2 | 5 | 40 | "There were lots of useful tools that veterans and first responders who have not had classes to talk with someone about their issues that it would be helpful for." – P3 |
| | Content not new for experienced users | 1 | 5 | 20 | "I have dealt with my stress a very long time ago, there was no new and helpful information on the educational content for me." – P7 |
| **Engagement Tools** | Breathing/music/videos perceived as useful | 3 | 5 | 60 | "When I was feeling stress the breathing and music would come in handy." – P3 |
| | Video-based learning preferred | 2 | 5 | 40 | "I used the videos a lot. I learn better listening and seeing." – P15 |
| | Journaling used infrequently | 1 | 5 | 20 | "I only wrote in the journal a few times." – P4<br>"I wish the breathing exercises had audio so you didn't have to look at the watch." – P15 |

| | | | |
|---|---|---|---|
| Desire for audio-guided features | 1 | 5 | 20 |

## 5. DISCUSSION

The present study evaluated a wearable-integrated digital mental health system embedded within an endurance cycling intervention for veterans, with the goal of characterizing longitudinal hyperarousal trajectories, symptom outcomes, and human-centered performance of the detection and self-management system in a naturalistic setting. Using generalized additive mixed models (GAMMs), we examined nonlinear temporal dynamics of hyperarousal (Moments of Stress) and self-reported symptom measures (GAD-7, PCL-5, PHQ-8) across three conditions.

### 5.1. Hyperarousal Trajectories

Baseline-normalized hyperarousal trajectories (derived from the number of MS confirmed by participants) differed significantly between study conditions, with the digital + physical intervention group exhibiting structured nonlinear dynamics that contrasted with the late-study escalation observed in the physical intervention-only condition. This pattern suggests that the digital intervention may help stabilize physiological stress responses following the endurance event.

From a behavioral perspective, the physical intervention alone provides strong physiological activation and social engagement, but without digital support, participants may lack mechanisms to interpret and regulate subsequent hyperarousal. One plausible explanation for the observed improvements in hyperarousal and symptom trajectories is that the digital intervention effectively closed a feedback loop between physiological detection and conscious awareness. The system provided real-time signals about hyperarousal states that may otherwise remain implicit, allowing participants to recognize and respond to stress in the moment. This mechanism aligns with the control theory of self-regulation, which posits that behavior change is facilitated when individuals receive timely feedback about internal states and can adjust actions accordingly (Carver & Scheier, 1982). This interpretation is consistent with prior wearable-based stress interventions showing that monitoring is most effective when paired with real-time behavioral support (Fuhrmann et al., 2025; Smith et al., 2020).

Importantly, substantial between-participant variability in baseline hyperarousal highlights the individualized nature of physiological stress signatures. Such heterogeneity reinforces the need for adaptive, user-centered modeling approaches in wearable mental health systems.

### 5.2. Psychological Symptom Trajectories

The digital + physical intervention group achieved its largest improvements in psychological symptom trajectories during the mid-study period following the endurance event, with partial rebound thereafter. For anxiety (GAD-7), the digital + physical intervention group demonstrated a clinically meaningful decrease in anxiety from ‘Mild’ (6.33) to ‘Minimal’ (1.67) levels of anxiety from baseline to Week 4; however, this reduction was not sustained, with a return back to ‘Mild’ levels of anxiety (4.40) by the final week.  Depressive symptoms (PHQ-8) followed a similar pattern, with an initial reduction in depressive symptoms from ‘Mild’ (6.75) to ‘Minimal’(2.25) depression from baseline to Week 3 – a change that did not quite reaching the estimated reliable change index of 5.2 points for PHQ-8 (Gyani et al., 2013) - with a return to ‘Mild’ depression (5.60) at follow-up. PTSD severity (PCL-5) decreased from 16.0 to a minimum of 7.83 mid-study but returned to 14.0 by Week 6, fluctuations that do not meet the estimated reliable change index of 18 (Marx et al., 2022). These patterns indicate acute improvement with partial loss of gains over time. However, it is also important to acknowledge that baseline levels of anxiety, depression, and PTSD for the digital + physical intervention group were below clinical thresholds, and therefore, it is difficult to suggest that these trajectories represent a “rebounding” effect, given that mean scores at follow-up also do not reach clinical significance.

The physical intervention-only group exhibited substantially higher baseline severity across all three psychological symptom measures compared to the digital + physical and at-home control conditions, indicating greater room for improvement at study entry. For anxiety (GAD-7), the physical intervention-only group began in the ‘Severe’ range (21.0) and showed a marked reduction during the endurance event to ‘Moderate’ anxiety levels (9.5) by Week 3. However, this improvement was not sustained, with anxiety returning to ‘Severe’ levels immediately after the event (21.0) at Week 4, before decreasing again to ‘Moderate’ levels (12.5) by Week 6. Depressive symptoms (PHQ-8) followed a similar trajectory, decreasing from “Severe” depression at baseline (24.0) to ‘Moderate’ levels (12.5) during the endurance challenge (Week 3), followed by a rebound to ‘Severe’ depression (23.5) at Week 4, and a subsequent reduction to ‘Moderately Severe’ depression (15.0) by Week 6. PTSD severity (PCL-5) also decreased sharply during the endurance event from 80.0 at baseline to 39.0 at Week 3 (above the reliable change index of 18 points and nearing the clinical threshold for probable PTSD at 32 points), before returning to 80.0 at Week 4 and remaining elevated at 62.67 by Week 6, well above commonly used clinical thresholds for probable PTSD (Marx et al., 2022; Weathers et al., 2013). These findings indicate that the endurance cycling intervention produced substantial short-term reductions in psychological distress, but these improvements were not sustained in the weeks immediately following the event. However, it is likely that self-regulation occurred to manage symptom severity in the weeks prior to follow-up. Importantly, the magnitude of these changes should be interpreted in the context of the group’s higher baseline severity, which allowed greater absolute reductions relative to the other conditions.

The at-home control group demonstrated more gradual and stable changes in psychological symptoms across the study period. For anxiety (GAD-7), mean scores declined from ‘Moderate’ anxiety at baseline (11.0) to ‘Mild’ levels by Week 6 (7.0). Depressive symptoms (PHQ-8) followed a similar trajectory, decreasing from ‘Moderate’ depression (11.0) to ‘Mild’ levels at follow-up (6.67). PTSD severity (PCL-5) also declined steadily over time, from 32.0 at baseline, a score consistent with the commonly used threshold for probable PTSD, to 17.33 by Week 6, falling below this threshold. Compared to the cycling conditions, these improvements occurred more gradually and without the sharp mid-study reductions observed during the endurance event. This pattern suggests that the observed symptom changes in the at-home control group may reflect the influence of ongoing self-monitoring and repeated psychological assessment rather than the acute physiological effects associated with intensive physical activity.

The absence of a significant linear time-by-condition interaction indicates that symptom trajectories were not well characterized by simple linear increases or decreases across the study period. Instead, the GAMM analysis revealed condition-specific nonlinear temporal patterns. Significant nonlinear time effects were detected only in the digital + physical intervention condition for anxiety (GAD-7) and depressive symptoms (PHQ-8), suggesting that symptom changes in this group followed a curved trajectory rather than a consistent linear trend over time. In contrast, the physical intervention-only and at-home control conditions did not show statistically reliable nonlinear time effects for these outcomes. For PTSD severity (PCL-5), the primary pattern reflected stable between-group differences, with the physical intervention-only condition exhibiting consistently higher scores across the study period compared to the other two conditions. Together, these results suggest that symptom change during the study was characterized more by dynamic temporal fluctuations within conditions than by clear linear treatment effects across groups.

The observed improvements in symptom trajectories likely reflect the interaction of two complementary mechanisms: physiological benefits associated with structured physical activity and cognitive-regulatory support provided by the digital intervention. Various research demonstrates that aerobic exercise can produce acute reductions in anxiety and depressive symptoms through neurobiological pathways such as endorphin release and autonomic regulation (Basso & Suzuki, 2016; Matta Mello Portugal et al., 2013). The symptom decreases observed during the endurance challenge in both physical intervention groups are consistent with these established exercise effects. However, the post-event trajectories differed across the two cycling conditions. The physical intervention-only group exhibited larger acute reductions accompanied by more pronounced post-event fluctuations, whereas the digital + physical intervention group showed more moderate changes with partial rebound, but with symptom levels generally remaining below baseline. The digital intervention provided additional scaffolding that may have supported consolidation of short-term improvements into more stable patterns of regulation. Real-time detection increased situational awareness of stress states, and immediate access to coping tools allowed behavioral strategies to be delivered at moments of heightened vulnerability.

Repeated cycles of detection and successful coping may strengthen participants' confidence in their ability to regulate distress (Aspinwall & Taylor, 1997).

The gradual symptom decline observed in the at-home control group likely reflects a combination of factors. Even without the structured endurance intervention and full digital intervention, participants engaged in continuous monitoring and weekly self-assessment, processes that may have promoted awareness and slight behavioral adjustment. At-home participants may have also been cycling on their own at home, but without a social, structured event like the Texas Challenge. Prior work on self-monitoring and ecological assessment suggests that repeated reflection on symptoms can function as a low-intensity intervention by increasing insight into emotional patterns and encouraging adaptive coping (Bennett-Levy et al., 2010; Hume, 2017).

### 5.3. Perceived Precision

By aggregating perceived precision to weekly means and aligning these values with weekly symptom assessments, we observed that higher symptom severity was consistently associated with greater confirmation of system-detected hyperarousal events. The results highlight the importance of human factors in the deployment of ML-enabled mental health systems. Perceived precision varied systematically with symptom severity and evolved over time, indicating that user trust is not static but develops through repeated interaction. This result is consistent with other work that has shown human operators' trust in automation systems increases over time as true positives are detected (Humr et al., 2023; X. J. Yang et al., 2017).

Across the three experimental conditions, distinct patterns emerged that help contextualize both symptom trajectories and perceived precision. The digital + physical intervention group consistently fell within the lower severity ranges for anxiety and depression across most weekly assessments. In contrast, the physical intervention-only group remained in higher severity ranges throughout the study despite transient improvements surrounding the endurance event. The at-home control group exhibited intermediate severity levels and greater variability, alternating between mild and moderate ranges across weeks.

These condition-level differences are important for interpreting perceived precision results. Participants in the digital + physical intervention group, who more frequently occupied lower symptom severity states, may have experienced fewer high-intensity hyperarousal episodes, contributing to different patterns of alert confirmation compared to the physical intervention-only group. Conversely, the consistently higher severity observed in the physical intervention-only group likely increased exposure to salient hyperarousal experiences. It is possible that individuals experiencing greater weekly symptom burden may be more attuned to physiological arousal states, resulting in stronger alignment between algorithmic detection and subjective experience. Alternatively, increased symptom severity may amplify bodily cues that the multimodal model captures more reliably. The mixed severity profile of the at-home control group provides an intermediate case that mirrors the variability observed in perceived precision trajectories. This work demonstrates that perceived precision covaries with symptom severity at a weekly timescale.

### 5.4. Veteran Perspectives and Takeaways

Thematic analysis further illustrates how veterans interpret and engage with real-time detection prompts during daily life. Overall, participants described the system as supportive for increasing awareness and prompting brief self-reflection after alerts. At the same time, perceived accuracy and alert acceptability varied across users and contexts, including reports that notifications could feel overly sensitive in non-stressful situations. This between-user variability is consistent with the highly idiosyncratic nature of hyperarousal experiences and supports the need for calibration to individual baselines and context-aware personalization to improve alignment between model outputs and user interpretations (Sadeghi et al., 2022).

From the perspective of these results, future AI/ML system design should underscore: (1) within-person model adaptation, including the establishment of personal baselines and dynamic thresholding; (2)

integration of contextual information, such as activity state and situational cues, to distinguish between different types of arousal, e.g., adaptive arousal (e.g., exercise-induced) vs ineffective arousal that leads to negative emotions, such as anger; (3) provision of explainability and transparency at the point of use, for instance, through concise cues that clarify the rationale for each alert; and (4) minimization of interaction burden (by minimizing excessive interaction, and so reducing the workload) to support sustained usability and trust in both daily and professional environments.

### 5.5. Strengths, Limitations and Future Directions

This study advances methodological contributions by integrating multimodal wearable features to improve real-time stress detection. By embedding these sensing capabilities within a veteran-centered self-management platform, this work moves beyond detection validation to examine intervention-level impact, representing one of the first naturalistic evaluations of a closed-loop, wearable-enabled digital mental health system for PTSD. Another key contribution of the study is the explicit treatment of inter-individual heterogeneity in hyperarousal experiences and detection perceptions as a meaningful signal, rather than as nuisance variance to be controlled or ignored. Through the integration of continuous multimodal sensing, real-time user confirmation, and evidence from naturalistic user experiences, this work systematically reveals between-person differences in baseline physiological states, contextual triggers, and individual tolerance for false alarms. Adopting this human-centered data science approach (Aragon et al., 2022) provides a foundation for future requirements in human-centered AI (Russell, 2019; Shneiderman, 2022) and machine learning (Fiebrink & Gillies, 2018) for wearable mental health systems.

Several limitations of the study warrant consideration. The small pilot sample size limits statistical power and generalizability. Symptom outcomes rely on self-report and may be influenced by response bias. Weekly aggregation of perceived precision, while appropriate for aligning with clinical assessments, may obscure finer-grained temporal dynamics. Use of digital intervention features was not standardized, preventing attribution of effects to specific components. Additionally, the perceived precision analyses involve a newly implemented multimodal algorithm integrating heart rate and accelerometer features; variability in user experience may partly reflect early-stage model calibration.

An additional limitation is the imbalance in baseline symptom severity across experimental conditions, which complicates direct comparison of improvement trajectories. Participants in the physical intervention-only condition reported substantially higher initial scores across all three psychological measures (GAD-7, PHQ-8, and PCL-5) compared to the other study conditions. Given the small sample size of this pilot study (n = 3 in the physical intervention-only group), random allocation did not produce balanced baseline symptom levels across groups. As a result, participants in this condition had greater potential for absolute symptom reductions relative to the other groups, and comparisons of symptom trajectories should therefore be interpreted cautiously. Although the study design allows examination of differential patterns over time, future work with stratified randomization and larger samples will be necessary to distinguish intervention effects from baseline differences.

Another limitation concerns the intervention structure. The digital self-management platform was deployed in conjunction with participation in an endurance cycling program that itself has documented psychological benefits in veteran populations. Although the inclusion of a physical-activity–only comparison group partially addresses this issue, the study design does not allow the unique contribution of the digital intervention to be fully isolated from the effects of the cycling program. As a result, observed improvements may reflect the combined influence of structured physical activity, peer engagement, and digital self-management support rather than the digital intervention alone. Future studies should evaluate the platform in additional contexts that allow clearer attribution of effects, such as deployments among veterans not participating in organized physical activity programs or trials.

Attrition in the present pilot trial was higher than anticipated. Several factors likely contributed to this pattern, including logistical challenges associated with deploying wearable devices during a multi-day endurance cycling event, variability in participant familiarity with smartwatch technology, and incomplete follow-up survey completion during the post-event monitoring phase.

Future protocols should incorporate proactive strategies to reduce participant burden and improve data completeness. These may include structured device onboarding sessions prior to the intervention, real-time technical support for synchronization issues, automated reminders for ecological momentary assessments and weekly surveys, and redundant data capture mechanisms to mitigate temporary device failures. Additionally, implementing staged run-in periods or pilot device testing before the start of the intervention may help identify technical issues and improve participant comfort with the monitoring system.

Future work should prioritize larger, longer-term studies examining adaptive personalization and user-in-the-loop model refinement. Another important direction for future work is to examine participant engagement with specific components of the digital intervention and their relationship to clinical outcomes. The present study treated the intervention as a unified system, but different features, such as breathing exercises, journaling tools, educational content, and responses to detection prompts, may contribute differentially to symptom change. Additionally, incorporating customizable notification strategies and contextual awareness may further enhance usability and effectiveness.

## 6. CONCLUSION

The present findings support the potential for wearable-integrated, human-centered digital self-management approaches to serve as scalable adjuncts to group-based physical activity interventions for veterans, with particular promise for sustaining benefits during the recovery phase following intensive activities. Notably, individual differences in perceived precision and alert acceptability were substantial, underscoring that heterogeneity among users should be regarded as a fundamental design consideration rather than as nuisance variance, especially in the development of closed-loop mHealth systems. In practical terms, future deployments of FWD-style systems should emphasize calibration to individual baselines and adaptive thresholds, incorporate context-aware detection mechanisms to distinguish between exercise-induced arousal and maladaptive hyperarousal, provide low-burden, just-in-time post-alert interventions such as rapid grounding or breathing techniques, and include lightweight transparency features to maintain user trust. Subsequent research should aim to replicate these findings in larger, sufficiently powered studies, systematically relate outcomes to feature-use data and contextual variables, and evaluate personalization strategies that minimize false-alarm rates while maintaining sensitivity to clinically significant hyperarousal.

## ACKNOWLEDGEMENTS

The authors thank Jacob M. Kolman, MA, ISMPP CMPP, senior technologist at Texas A&M University, for critical review and editing on an earlier version of this manuscript.

**CRediT authorship contribution statement**

**Alan Ta:** Investigation, Methodology, Data Curation, Formal analysis, Visualization, Writing - Original Draft **Nilsu Salgin:** Formal analysis, Writing - Original Draft **Mustafa Demir, PhD:** Investigation, Methodology, Data Curation, Formal analysis, Visualization, Writing - Original Draft **Kala Phillips Reindel, PhD:** Methodology, Writing- Reviewing and Editing **Farzan Sasangohar, PhD:** Conceptualization, Supervision, Methodology, Writing- Reviewing and Editing, Funding acquisition

**Declaration of competing interest**

The authors declare that they have no known competing financial interests or personal relationships that could have appeared to influence the work reported in this paper.

**Funding**

This research did not receive any specific grant from funding agencies in the public, commercial, or not-for-profit sectors.

**Declaration of generative AI and AI-assisted technologies in the manuscript preparation process**

During the preparation of this work, the authors used ChatGPT to assist with generating and refining code for data visualization used in the analysis workflow. After using this tool, the authors reviewed, tested, and edited the code as needed and take full responsibility for the accuracy and integrity of the analyses and the content of the published article.

**Consent and Approval**

Written informed consent was obtained from each study participant prior to onboarding. This study was approved by the Texas A&M University Institutional Review Board (STUDY2025-0124) on April 4th, 2025. This study was registered and approved on Clinicaltrials.gov (NCT06993012).

**Data Availability**

De-identified participant data underlying the findings of this study, along with a data dictionary and statistical analysis scripts used for the reported analyses, may be made available upon reasonable request to the corresponding author, subject to approval by the Texas A&M University Institutional Review Board and data use agreements designed to protect participant confidentiality. Due to intellectual property considerations associated with an active patent application, the full source code of the machine learning algorithm used for hyperarousal detection cannot be publicly released at this time.